\documentclass[]{IEEEtran}

\usepackage{xcolor}
\usepackage{graphicx}
\usepackage{amsfonts} 
\usepackage{amsmath}
\usepackage{amsthm}
\usepackage{bbm}
\usepackage{hyperref}

\newtheorem{lemma}{Lemma}
\newtheorem{definition}{Definition}

\title{A Frequency-Structure Approach for \\ Link Stream Analysis}
\author{\IEEEauthorblockN{Esteban Bautista and Matthieu Latapy} \\
\IEEEauthorblockA{
Sorbonne Université, CNRS, LIP6, F-75005 Paris, France \\
\vspace*{-0.8cm}}
}

\begin{document}

\maketitle
\thispagestyle{plain}
\pagestyle{plain}

\begin{abstract}
A link stream is a set of triplets $(t, u, v)$ indicating that $u$ and $v$ interacted at time $t$. Link streams model numerous datasets and their proper study is crucial in many applications. In practice, raw link streams are often aggregated or transformed into time series or graphs where decisions are made. Yet, it remains unclear how the dynamical and structural information of a raw link stream carries into the transformed object. This work shows that it is possible to shed light into this question by studying link streams via algebraically linear graph and signal operators, for which we introduce a novel linear matrix framework for the analysis of link streams. We show that, due to their linearity, most methods in signal processing can be easily adopted by our framework to analyze the time/frequency information of link streams. However, the availability of linear graph methods to analyze relational/structural information is limited. We address this limitation by developing (i) a new basis for graphs that allow us to decompose them into structures at different resolution levels; and (ii) filters for graphs that allow us to change their structural information in a controlled manner. By plugging-in these developments and their time-domain counterpart into our framework, we are able to (i) obtain a new basis for link streams that allow us to represent them in a frequency-structure domain; and (ii) show that many interesting transformations to link streams, like the aggregation of interactions or their embedding into a euclidean space, can be seen as simple filters in our frequency-structure domain. 

\end{abstract}

\section{Introduction}
\label{Sec.Introduction}

Financial transactions, phone calls, or network traffic are examples of data that can be very well modeled as a link stream \cite{latapy2018stream, Latapy2019WeightedBO}: a set of possibly weighted triplets $(t, u, v)$ indicating that $u$ and $v$ interacted at time $t$. For example, a triplet can model that a bank account $u$ made a transaction to an account $v$ at a time $t$, or that a computer $u$ sent a packet to a computer $v$ at a time $t$ (weights may represent amounts or sizes). Link streams have gained considerable attention in recent years as numerous phenomena of crucial interest such as financial frauds, network attacks, or fake news, correspond in them to clusters of interactions with some distinctive dynamical and structural signature. For this reason, the development of techniques allowing a precise understanding of the dynamical and structural properties of link streams has become a subject of utmost importance. 

Traditionally, link streams are seen as collections of time series (one for each relation $u, v$) or as sequences of graphs (one for each time $t$). These interpretations then allow to use the frameworks of signal processing and graph theory to study their dynamical and structural information, respectively. However, the very sparse nature of most real-world link streams causes these approaches to be frequently inconclusive. Namely, the graphs from the sequence often contain so few edges that it is hard to extract useful structural information from them, while the time series of relations are often too spiky to allow the extraction of insightful dynamical patterns. For these reasons, it is a very common practice to transform a sparse link stream into a denser one by aggregating all interactions contained within pre-defined time windows, which results in a new link stream with properties that are simpler to study. Yet, this comes at the price of potentially no longer containing the information necessary to detect an event of interest. For example, if two links that appear in succession is key to spot an event, then the aggregation process is likely to destroy this information. 

Interestingly, aggregation is not the only situation where the above problem may occur, as the signal and graph interpretations are often used to transform link streams into new objects where decisions are made. Such new objects may be another link stream \cite{chiappori2021quantitative, ribeiro2013quantifying, paranjape2017motifs}, a time series whose samples summarize structural information \cite{Fontugne2017ScalingII, ozcan2017supervised, bhatia2020midas, kodali2020value}, or a graph whose edges summarize dynamical information \cite{peng2019frequency, zhang2019wavelet, Chang2021FFADEFF}. In all these situations, information about the raw link stream is potentially destroyed during the transformation processes, making it pertinent to raise the question: what dynamical and structural information of the raw link stream carry into the new object? Addressing this question is the fueling force of this work. 

This work shows that, in order to address the aforementioned question, it is necessary that operators from signal processing and graph theory can be algebraically interchanged, which can be achieved by restricting to the subset of linear signal and graph operators. Based on this insight, we introduce a new linear framework for the analysis of link streams. In this framework, we represent a link stream by a simple matrix whose and its processing amounts to multiply the link stream by other matrices that encode for linear graph and signal operators. Notably, the linearity of most signal processing methods implies that they can be readily incorporated into our framework as a means to process the time properties of link streams. However, the processing of the relational properties poses a challenge as the availability of linear graph operators is scarce, raising the question of how to design meaningful graph methods that satisfy linearity. 

To address the aforementioned challenge, we interpret graphs as functions and then adapt signal processing methods to process such functions. In particular, we leverage this methodology to (i) develop a new basis for graphs that allow us to represent them in a structural domain; and (ii) develop filters for graphs that allow us to suppress structural information from them in a controlled manner. These results can therefore be seen as graph analogs of the Fourier/wavelet transform and of frequency filters. By combining such developments with their time domain counterpart via our framework, we are able to (i) develop a new basis for link streams that allow us to represent them in a frequency-structure domain; and (ii) address our motivating question by showing that transformations to link streams, like aggregation, can be seen as simple filters in our frequency-structure domain. Indeed, we show that our results may have other interesting applications, like the extraction of patterns with specific structure and frequency signatures and the quantification of the data regularity which paves the way to do machine learning directly on link streams.

\section{Definitions and problem statement}
\label{Sec.Problem_Statement}

\subsection{Definitions}
We formally define a link stream as a quadruplet $\mathcal{L} = (T, V, D, L)$, where $T$ is a set of times, $V$ is a set of vertices, $D \subseteq T \times V \times V$ is the set of link stream triplets, and $L : T \times V \times V \to \mathbb{R}$ is a function associating a weight to triplets, so that $L(t, u, v) = 0$ if $(t, u, v) \notin D$ and, for the case of unweighted link streams, $L(t, u, v) = 1$ if $(t, u, v) \in D$. We restrict to discrete-time link streams, thus we set $T = \mathbb{Z}$. The infinite size of $T$ is assumed for theoretical simplicity, when processing link streams numerically $T$ is restricted to a bounded interval. 

We denote the space of all possible relations between the vertices of the link stream by $\mathcal{E} = V \times V$. Relations are considered directed, hence $(u, v) \neq (v, u)$. We also assume the members of $\mathcal{E}$ to be indexed so that $e_k \in \mathcal{E}$ refers to its $k$-th element. The index can be assigned arbitrarily, Section \ref{Sec.Graph_Decomp_Partitioning} covers techniques that re-index the elements in useful manners. We denote $|\mathcal{E}| = M$ and assume, without loss of generality, that $|\mathcal{E}|$ and $|V|$ are powers of two. Subsets $\mathcal{E}' \subseteq \mathcal{E}$ are systematically referred to as structures or motifs since we assume they have some structural significance: they are made of spatially close relations that may form cliques, communities, stars, etc. 

The time series associated to $e_k$ is denoted as $e_k(t)$, where $e_k(t) = L(t, e_k)$. The graph associated to time $t$ is denoted as $G_t = (V, E_t, f_{G_t})$, where $E_t = \{ e_k \in \mathcal{E} : (t, e_k) \in D \} $ refers to its set of edges and $f_{G_t} : \mathcal{E} \to \mathbb{R}$ refers to a weight function defined as $f_{G_t}(e_k) = L(t, e_k)$. We recall that $\mathcal{E}$ refers to the set of all possible relations, while $E_t$ refers to only those that exist in $G_t$. Thus, we have that $E_t  \subseteq \mathcal{E}$. We consider unweighted graphs as graphs with unit-weights, thus we also include the weight function when referring to unweighted graphs.

Given two unweighted graphs $G_1 = (V_1, E_1, f_{G_1})$ and $G_2 = (V_2, E_2, f_{G_2})$, the distance of $G_1$ with respect to $G_2$ is given as $dist(G_1, G_2) = |E_1| - |E_1 \cap E_2|$, which counts the number of edges in $G_1$ that are not in $G_2$. The edit distance between $G_1$ and $G_2$ is given as $edit(G_1, G_2) = dist(G_1, G_2) + dist(G_2, G_1)$, which counts the number of different edges between $G_1$ and $G_2$.

\subsection{Problem statement}

As mentioned in the introduction, a link stream is frequently transformed into another link stream, a time series, or a graph. In this subsection, we give two motivating examples where such transformations are employed. The goal of the examples is to highlight that, in order to understand how the dynamics and structure of a link stream carry into the new object, signal and graph operators must be represented in the latent space of each other, which may be hard due to their different nature. Then, we show that one solution to this problem is by restricting to linear graph and signal operators. For our discussion, we study the link stream through its weight function, which we recall is a two-variable function of time and relations, i.e., $L(t,e)$. Moreover, we denote signal and graph operators by $s$ and $g$, respectively. We stress that a signal operator, when applied to $L(t,e)$, acts on all $t$ with $e$ fixed and the graph operator does the converse. 
\begin{figure}[t]
\center
  \includegraphics[width=0.48\textwidth]{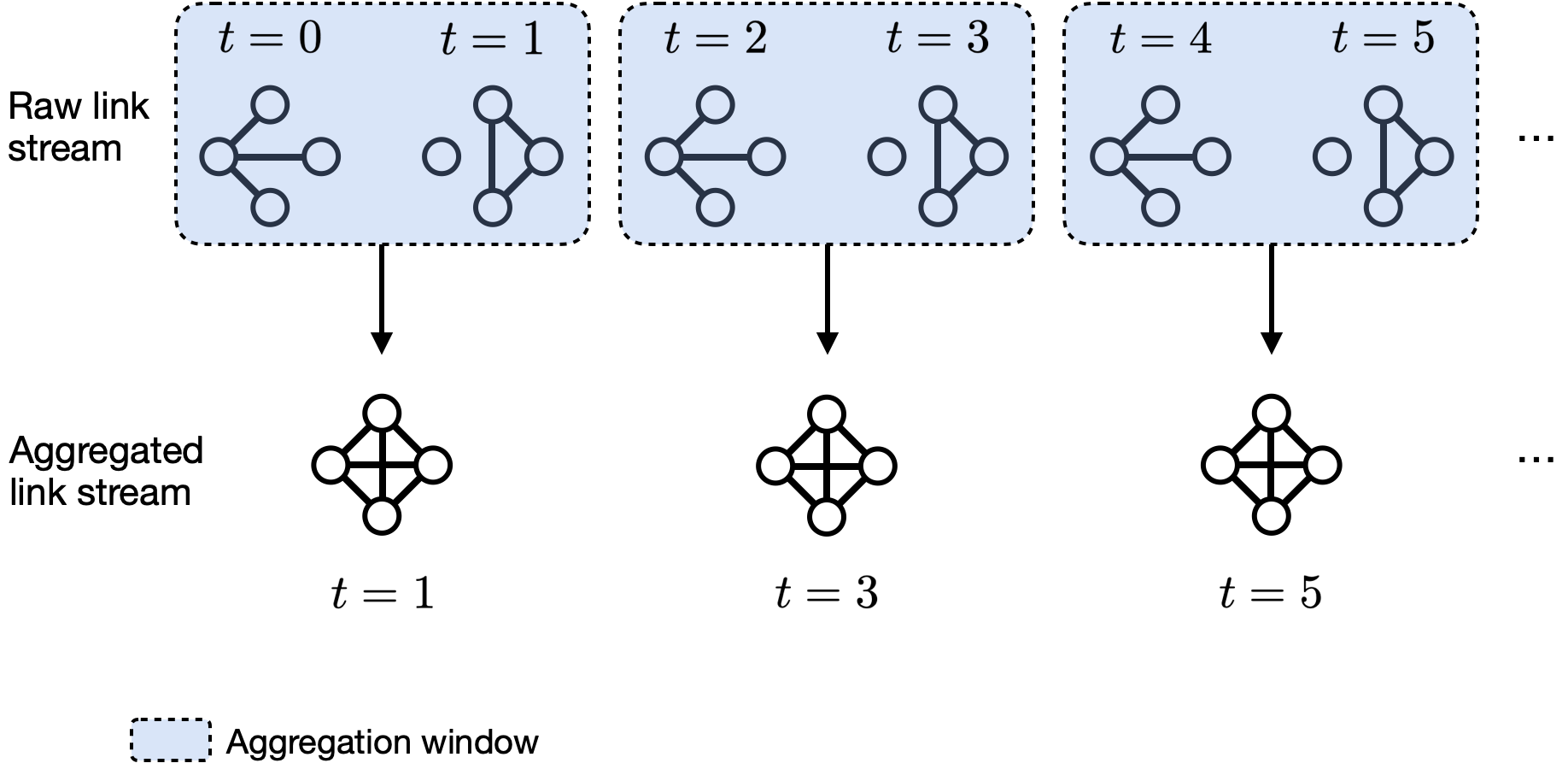}
  \caption{Illustration of the aggregation of interactions in link streams. A four-vertex raw link stream consisting of out-of-phase claw and triangle structures is aggregated via a 2-sample window that makes emerge a constant clique.}\label{fig.sec.2.Agg_LS}
\end{figure}

{\bf Motivating example (aggregation).}~A very common example where a signal operator is used to transform a link stream is the aggregation of interactions, which is a process that can drastically change the properties of a link stream. To illustrate this, consider the example of Figure \ref{fig.sec.2.Agg_LS}, where we show a link stream consisting of alternating triangle and claw graphs (i.e., they oscillate at frequency $1/2$). Then, this link stream is aggregated by means of a 2-sample window, resulting in a new link stream that consists of a constant clique (i.e., it has zero frequency). Hence, the aggregation process makes the oscillating claw and triangle disappear in order to make emerge a new structure and a frequency that were not initially in the link stream, highlighting the importance of understanding how the properties of a link stream change when it is aggregated. 

Mathematically, the aggregation process can be modeled as an operator $s$ that adds interactions in a sliding window as
\begin{equation}\label{Eq.sec.2.aggregation}
	\hat{L}(t, e) = s(L(t,e)) = \sum_{k = 0}^{K - 1} L(t - k, e). 
\end{equation}
where $\hat{L}$ refers to the aggregated link stream. To faithfully model the process of Figure \ref{fig.sec.2.Agg_LS}, a sub-sampler should also be applied to $\hat{L}$ in order to only retain aggregates from non-overlapping windows. Sub-sampling $\hat{L}$ amounts to simply dilate its frequency spectrum. Thus, we omit it for simplicity and focus on the real challenge which is to relate the properties of $L$ with those of $\hat{L}$. 

We begin by relating the dynamical information of $L$ and $\hat{L}$, which can be done by relating their frequency content. Since $s$ is a linear time invariant operator, the convolution theorem indicates that the frequency content of $\hat{L}$ is that of $L$ multiplied by the frequency response of the operator $s$. Thus, since $s$ acts on the time axis, its impact on the frequency information is straightforward to assess with standard signal processing results.  

A more difficult task is, however, to relate how the structural information of $\hat{L}$ relates to that of $L$. To show this, let us assume that we dispose of an operator $g$ that measures some structural property of the graphs from the sequence. For example, $g$ may be an operator that counts their number of triangles. In this case, if we aim to know how $g(\hat{L}(t,e))$ relates to $g(L(t,e))$, then it is necessary to know the equivalent of the aggregation operator $s$ in the latent space of the $g$ operator. This is, we must find an alternative representation of $s$, denoted $s'$, that transforms the number of triangles in the graphs of $L$ into the number of triangles in the graphs of $\hat{L}$, so that $g(\hat{L}(t,e)) = s'( g( L(t,e) ) )$. Due to their different nature, finding the equivalent representation of $s$ in the latent space of $g$ is a difficult task. 

One possible solution to the above problem consists in looking for graph operators $g$ that commute with $s$, as $g(\hat{L}(t,e)) = g(s(L(t,e))$. Notably, this can be achieved by restricting to linear graph operators of the form $g(L(t,e))(t,i) = \sum_e q(i, e) L(t,e)$, where $i$ is a dummy index that may refer to a coordinate (if $g$ transforms a graph into a feature vector) or to an edge (if $g$ transforms the graph into another graph). In this case, the linearity of $g$ and $s$ nicely combine to obtain
\begin{align}
g( \hat{L}(t,e) ) &= \sum_{e} q(i,e) \hat{L}(t,e)   \\
&=  \sum_{e} q(i, e) \sum_{k = 0}^{K} L(t - k, e) \\
&=  \sum_{k = 0}^{K} \sum_{e} q(i, e)  L(t - k, e) \\
&= \sum_{k = 0}^{K} g(L(t - k, e) ),
\end{align}
which expresses the structural information of $\hat{L}$ as a combination of the structural information of $L$. 

{\bf Motivating example (embedding).}~A popular approach to transform a link stream into a time series is by means of graph embedding methods. Graph embedding refers to the process of mapping a graph to a scalar or a point in an euclidean space, so that structurally similar graphs map to close points in the embedded space. These techniques, which are commonly based on neural networks \cite{yu2018netwalk, ijcai2018p505, pareja2020evolvegcn}, spectral methods \cite{li2017attributed, zhu2018high}, or diffusion processes \cite{du2018dynamic, nguyen2018continuous}, are hence applied to the graph sequence in order to obtain a time series (or a group of them) that reflects the structural evolution of the link stream. Since the dynamics of such time series are usually studied to look for anomalies \cite{mahdavi2018dynnode2vec, zheng2019addgraph} or events \cite{kumar2019predicting, Know-evolve2017} in the link stream, it is natural to ask to what extend the dynamics of such time series reflect the dynamics of the original link stream.  

In oder to address this question, let us assume that that the graph embedding method is represented by the operator $g$ so that the embedding into the $i$-th time series is represented as $\hat{L}(t, i) = g(L(t,e))$. Then, our goal is to relate the frequency content of $\hat{L}$ with that of $L$. For this, let us assume that $s$ refers to the Fourier transform operator. In this case, if we aim to relate $s(\hat{L}(t,i))$ with $s(L(t,e))$, then we run into a similar challenge as in our previous example: it is necessary to know the equivalent of the graph embedding in the Fourier transform domain, which is unclear to know. Nonetheless, as in our previous example, we stress that if the embedding method was linear, then its linearity combines well with the linearity of the Fourier transform to obtain
\begin{align}
s(\hat{L}(t)) &= \sum_{t = -\infty}^{\infty} \hat{L}(t) e^{-j \omega t} \\
&= \sum_{t = -\infty}^{\infty} \left[ \sum_{e} q(i, e) L(t,e) \right] e^{-j \omega t}  \\
&= \sum_e q(i,e) \sum_{t = -\infty}^{\infty} L(t,e) e^{-j \omega t} \\
&= \sum_e q(i,e) s(L(t,e))
\end{align}
which expresses the frequency content of $\hat{L}$ as a combination of the frequency content of $L$. 

From our two motivating examples, it can be seen that, as along as we stick to linear signal and graph operators, it is possible to characterize the impact of transformations to a link stream. This observation clearly opens the question of what signal and graph operators satisfying linearity allow a meaningful characterization of transformations like aggregation or embedding. The remainder of this paper investigates this problem. For this, we first introduce a linear matrix framework for link stream analysis (Section \ref{Sec.Matrix_Framework}). Then, we develop linear operators to analyze graphs (Section \ref{Sec.Graph_Decomp}). Lastly, we combine the new linear graph methods with signal processing ones to soundly represent transformations as simple filters in a frequeny-structure domain (Section \ref{Sec.LS_Decomp}).

\section{A linear framework for link stream analysis}
\label{Sec.Matrix_Framework}

In this section, we formalize our insights from Section \ref{Sec.Problem_Statement} by proposing a linear matrix framework for the processing of link streams. This matrix framework offers two main advantages: (i) it allows to unify the signal and graph points of view described above; and (ii) it allows to easily study the joint impact that operators (or combinations of them) have on the time and relational properties of link streams. To do this, we simply represent link streams and operators as matrices that get multipled. Namely, we represent a link stream $\mathcal{L}$ by the matrix $\mathbf{L} \in \mathbb{R}^{|T| \times |\mathcal{E}|}$ given as
\begin{equation}
\mathbf{L}_{t, k} = L(t, e_k).
\end{equation}
This matrix representation unifies the signal and graph interpretations, as the $k$-th column of $\mathbf{L}$ corresponds to $e_k(t)$ and the $t$-th row corresponds to the graph $f_{G_t}(e)$. This therefore implies that the application of any linear signal processing operator can be simply modelled as a matrix $\mathbf{H}$ that multiplies $\mathbf{L}$ from the left as 
\begin{equation}\label{Eq.sec.3.Hprod}
	\widehat{\mathbf{L}} = \mathbf{H} \mathbf{L},
\end{equation}
where the notation $\widehat{\mathbf{L}}$ represents a processed link stream. Similarly, the rows of $\mathbf{L}$ coding for the graph sequence implies that the application of any linear graph operator can be modeled as a matrix $\mathbf{Q}$ multiplying $\mathbf{L}$ from the right as 
\begin{equation}\label{Eq.sec.3.Qprod}
	\widehat{\mathbf{L}} = \mathbf{L} \mathbf{Q}.
\end{equation}
From equations (\ref{Eq.sec.3.Hprod}) and (\ref{Eq.sec.3.Qprod}) it can be clearly seen that (i) any chain of operators $\mathbf{H}_1 \dots \mathbf{H}_n$ or $\mathbf{Q}_1 \dots \mathbf{Q}_n$ can be reduced to an equivalent one that consists of their product; and (ii) signal and graph operators can be straightforwardly combined as
\begin{equation}\label{Eq.sec.2.Joint-Analysis}
	\mathbf{\widehat{L}} = \mathbf{H} \mathbf{L} \mathbf{Q}.
\end{equation}
Equation (\ref{Eq.sec.2.Joint-Analysis}) constitutes our proposed linear matrix framework for link stream analysis. Notice that it allows us to formalize our research question as the one of finding matrices $\mathbf{H}$ and $\mathbf{Q}$ which characterize or model interesting transformations to link streams. 

We stress that the field of signal processing has established a large number of powerful linear concepts to analyze signals, such as Fourier and wavelet transforms, convolutions, sampling theorems, correlations, or filters. Thus, due to their linearity, these methods can readily take the role of the matrix $\mathbf{H}$ and be used analyze the time properties of link streams. Yet, to analyze the relational properties, the availability of linear graph methods is scarce. Indeed, while various useful graph concepts like degree measures, edge counts, or cut metrics are linear, it is often necessary to rely on more advanced techniques based on motif/community searches, spectral decompositions, or graph embeddings. However, these more advanced approaches are adaptive and non-linear, meaning that they cannot be modeled as a single matrix $\mathbf{Q}$ that can be plugged into our framework. Section \ref{Sec.Graph_Decomp} aims to address this issue by developing new linear methods for graphs. Section \ref{Sec.LS_Decomp} then revisits our framework by incorporating such developments.

\section{Linear methods for graphs}\label{Sec.Linear-Graph-Methods}
\label{Sec.Graph_Decomp}

The goal of this section is to develop methods that can take the role of the matrix $\mathbf{Q}$ in (\ref{Eq.sec.2.Joint-Analysis}). We exploit the fact that signal processing offers a powerful set of linear methods to process functions. Thus, we approach the problem by interpreting graphs as functions and by adapting signal processing techniques to process such functions. From this perspective, the only difference between a time series and a graph is that the former is a function of the form $f_{s} : \mathbb{R} \to \mathbb{R}$ while the latter is of the form $f_{G} : \mathcal{E} \to \mathbb{R}$. The main challenge lies in adapting signal processing methods to take into account the non-ordered nature of the domain $\mathcal{E}$ of $f_{G}$. 

In particular, this section leverages the above approach to develop a structural decomposition for $f_{G}$ and thoroughly explores its implications. The motivation for developing a decomposition is to establish, for graphs, a tool that offers benefits that are similar to those offered by the Fourier transform in signal processing. Namely, by representing signals in a frequency domain, the Fourier transform allows to see that many transformations to and between signals amount to simply amplify or attenuate some of their frequency information. Thus, we aim an analog for graphs: to have a representation that permits to study what transformations to and between graphs can be seen as simple ways to emphasize or suppress some of their structures. 

The section is structured as follows. Section \ref{Sec.Graph_Decomp_Decomp} develops a multi-scale structural decomposition for graphs. Section \ref{Sec.Graph_Decomp_Partitioning} investigates two techniques to automatically construct the basis elements of the decomposition. Section \ref{Sec.Graph_Decomp_Embedding}, shows that the decomposition can be interpreted as a linear graph embedding method. Section \ref{Sec.Graph_Decomp_Filters} introduces structural filters for graphs. 
\begin{figure*}[t]
  \includegraphics[width=\textwidth]{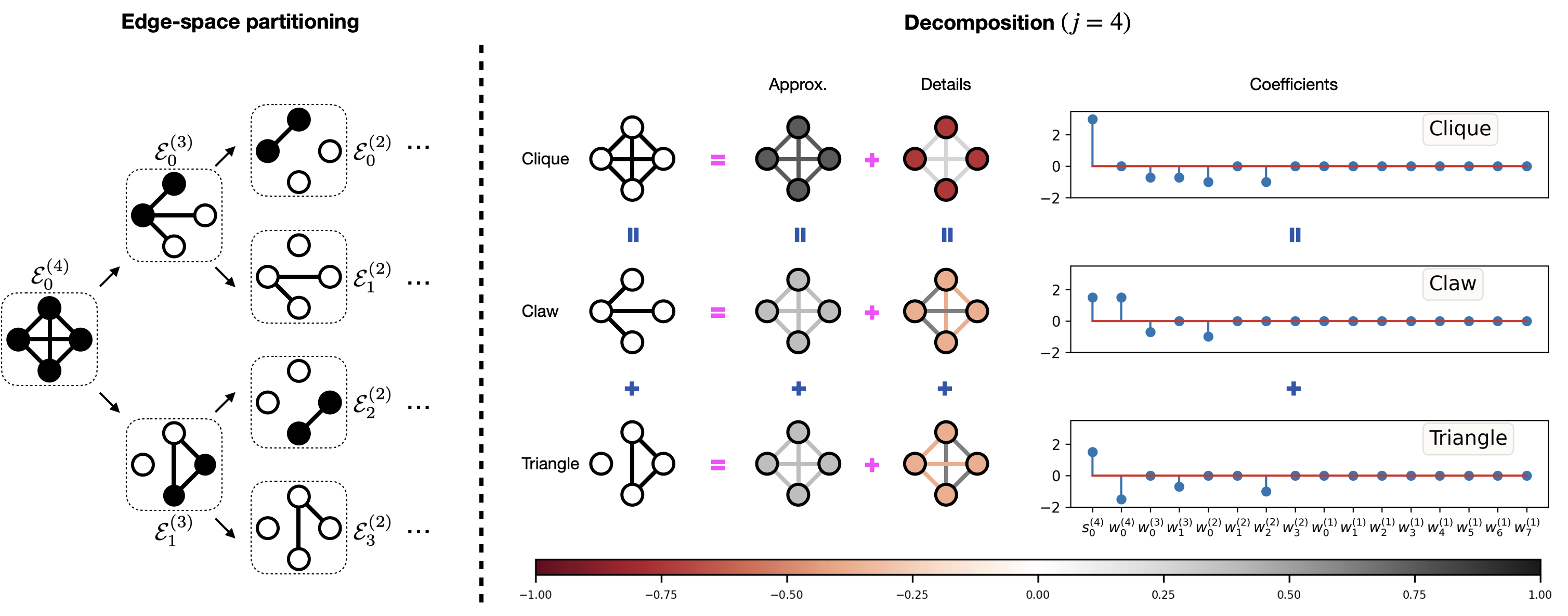}
  \caption{Illustration of the proposed graph decomposition on graphs with four vertices. The left panel displays the recursive partition of the relation-space. The right panel shows the application of the decomposition on the clique, claw and triangle graphs. Horizontally, the panel illustrates (i) how such graphs can be decomposed into coarse-grain and detailed parts; and (ii) their decomposition coefficients. Vertically, the panel shows how the aggregation process can also be studied in the decomposition domain. For the figure, edges are shown as undirected albeit the formalism considers such edges as two directed ones pointing in opposite directions. Colored nodes represent self-loops and colors encode edge-weights.} \label{Fig.sec.4.decomposition}
\end{figure*}

\subsection{A new decomposition for graphs}\label{Sec.Graph_Decomp_Decomp}
Our aim here is to develop a structural decomposition for $f_G$. In signal processing, a decomposition refers to the process of fixing a set of elementary signals and then expressing an arbitrary signal as a weighted combination of them. The weighting coefficients then allow to see the importance of an elementary signal in the decomposed signal. For example, in the Fourier case, the elementary signals are complex exponentials of varying frequencies, meaning that the Fourier transform coefficients reflect the importance of frequencies in the decomopsed signal. In our case, we aim to track the importance of structures in $f_{G}$, thus we search for a set of structurally-meaningful functions that allow us to express $f_{G}$ as a linear combination of them. 

To find such functions, we begin noticing that many interesting structures in graphs can be of different scales. For example, a structure of interest may consist of a large group of vertices forming a community, or it may consist of a small clique confined to the boundaries of the graph. Interestingly, if the vertices of such community or clique are referred to by the set $V_s \subset V$, then we have that, in both cases, the function $f_G$ is dense on the set $V_s \times V_s$. This is, for most edges $e' \in V_s \times V_s$ we have that $f_{G}(e') \neq 0$. If we further consider that non-zero edge weights are often very similar in magnitude (equal for unweighted graphs), then we can clearly see that the function $f_G$, on the set $V_s \times V_s$, can be very well approximated by a constant function also supported on the set $V_s \times V_s$. Thus, building from this observation, we propose to use constant functions supported on sets representing structures of interest as our group of elementary functions to decompose $f_{G}$. This is, we propose to do a multi-resolution analysis of $f_{G}$ by piece-wise constant-functions as our decomposition. 

In the signal processing literature, a decomposition of signals by piece-wise constant functions corresponds to the classical Haar wavelet transform. In it, a set of piece-wise constant functions (known as scaling functions) and a set of tripolar functions (known as wavelet functions) are constructed so that they form an orthornormal basis that can be used to expand any signal. Hence, our aim is to adapt Haar wavelets to the graph setting in a way that scaling and wavelet functions now acquire structural significance. Yet, we stress that this adaptation is not straightforward. In signals, the scaling and wavelet functions are built by shifting and dilating a primitive pulse-shaped signal. However, the notions of shifting and scaling are not well defined for functions supported on $\mathcal{E}$ due to its unordered nature. In the following, we demonstrate that we can still construct a Haar multi-resolution analysis for graphs by recursively partitioning the set $\mathcal{E}$. 

To begin, let us set $\mathcal{E}^{(\log_2(M)) }_0 = \mathcal{E} $ and recursively partition this set according to the following rule: $\mathcal{E}_k^{(j+1)} = \mathcal{E}_{2k}^{(j)} \cup \mathcal{E}_{2k+1}^{(j)}$, with $ \mathcal{E}_{2k}^{(j)} \cap \mathcal{E}_{2k+1}^{(j)} = \emptyset$ and $|\mathcal{E}^{(j)}_{2k}| = | \mathcal{E}^{(j)}_{2k +1}|$, until we obtain singletons $\mathcal{E}_{k}^{(0)} = e_k$. Then, based on this partitioning, we can define a set of scaling functions as:

\begin{equation}\label{Eq.sec.4.scaling_function}
\phi_k^{(j)}(e) = 
\begin{cases}
\sqrt{ 2^{-j} }  & e \in \mathcal{E}_{k}^{(j)} \\[5pt]
0 & \text{otherwise}
\end{cases} 
\end{equation}
and a set of wavelet functions as:
\begin{equation}\label{Eq.sec.4.wavelet_function}
\theta_k^{(j)}(e) = 
\begin{cases}
\sqrt{ 2^{-j}} & e \in \mathcal{E}_{2k}^{(j)} \\[5pt]
- \sqrt{ 2^{-j}} & e \in \mathcal{E}_{2k + 1}^{(j)} \\[5pt]
0 & \text{otherwise}
\end{cases} 
\end{equation}
It is easy to verify that our scaling and wavelet functions satisfy the following properties: (i) scaling and wavelet functions are pair-wise orthonormal; (ii) wavelet functions are pair-wise orthonormal; (iii) scaling functions, for a fixed $j$, are pair-wise orthonormal; and (iv) for level $j = \log_2(M) - \ell$, there are $2^{\ell}$ associated scaling functions and $2^{\ell}$ associated wavelet functions. These properties imply that, for a fixed $j$, the collection $\{ \phi_k^{(j)} \} \cup \{ \psi_k^{(\ell)} \}$, for all $k$ and $\ell \leq j$, forms a set of $M$ orthonormal functions that constitute a basis for functions supported on $\mathcal{E}$ (i.e., graphs). This basis, which constitutes an adaptation of the Haar multi-resolution analysis to graphs, allows us to decompose a graph in terms of the scaling and wavelet functions at different levels of resolution as indicated by $(j)$. This is done as follows: 
\begin{equation} \label{eq.sec.4.decomp_main}
f_{G}(e) = \sum_{k} s^{(j)}_k \phi^{(j)}_k(e) + \sum_{\ell \leq j} \sum_{k} w^{(\ell)}_k  \theta_k^{(\ell)}(e), 
\end{equation}
where 
\begin{equation}
s^{(j)}_k = \langle f_{G}, \phi^{(j)}_k \rangle
\end{equation}
refers to a scaling coefficient and 
\begin{equation}
w^{(j)}_k = \langle f_{G}, \theta^{(j)}_k \rangle 
\end{equation}
refers to a wavelet coefficient. Equation (\ref{eq.sec.4.decomp_main}) shows that a graph $f_{G}$ can be split in two main parts. On the one hand, the first term on the right-hand side constitutes a coarse grain approximation of $f_{G}$ at resolution level $(j)$. To see this, recall that scaling functions represent structures of interest at some resolution level, like the aforementioned community or clique. Thus, such first term constitutes the best possible approximation of $f_{G}$ given by such structures. The second term on the right-hand side contains, consequently, all the necessary details to recover $f_{G}$ from its coarse-grain approximation. Indeed, by simple algebra it can be shown that the wavelet coefficients at level $(\ell)$ contain the information to recover the coarse-grain approximation of $f_{G}$ at level $(\ell - 1)$ from the one at level $(\ell)$. 

{\bf Illustrating example (claw, triangle, clique).}~To better illustrate these concepts, let us give a practical example of our decomposition in Figure \ref{Fig.sec.4.decomposition}. For our example, we choose to work with the space of graphs defined on four vertices. This allow us to revisit our claw, triangle, and clique graphs discussed in Section \ref{Sec.Problem_Statement}. Since we aim to work with graphs of four vertices, the first step in our decomposition consists in partitioning the edge space associated to such graphs (which contains 16 directed edges, including self-loops). This partitioning procedure is depicted in the left panel of Figure \ref{Fig.sec.4.decomposition}. We begin with a single structure $\mathcal{E}_{0}^{(4)}$ which contains all the relation-space. This structure can thus be considered as a clique with self loops (as illustrated by the colored vertices). Then, we generate the structures at the next resolution level, this is $\mathcal{E}_{0}^{(3)}$ and $\mathcal{E}_{1}^{(3)}$. To do this, we split $\mathcal{E}_{0}^{(4)}$ in two equal-sized parts, which, for convenience, we pick as a claw and triangle with self-loops. These structures are then recursively partitioned until they cannot be further divided. 

Based on the structures found above, we can then define our scaling and basis functions according to (\ref{Eq.sec.4.scaling_function}) and (\ref{Eq.sec.4.wavelet_function}). Since they form a basis, we can use them to decompose any graph living in the space of graphs of four vertices, like the claw, triangle, and clique graphs shown in the right panel. For our example, we select to decompose them at resolution level $j = 4$, which is the coarsest one. This means that we will find the best approximations of such graphs by means of a clique with self-loops of constant weight (first term of (\ref{eq.sec.4.decomp_main})), where the edge weights are determined by the coefficient $s_{0}^{(4)}$. Naturally, the clique admits a very good approximation as its only difference with $\mathcal{E}_0^{(4)}$ lies on the self-loops, reason for which the edge-weights of the approximation graph are large and the detail graph concentrates its large-magnitude edges on the self-loops. Since the claw and triangle are structurally farther, their approximation graphs have less important weights and it is necessary to add to them an important amount of structural details to recover the initial graphs, reason for which their detailed graphs have important weights for most edges. 

Even though the claw and triangle approximations are not as accurate, their decomposition at this resolution level is still useful to better understand what occurs when such graphs are aggregated. Namely, due to the linearity of our decomposition, we have that the decomposition of the sum of two graphs is equal to the sum of their individual decompositions. This implies that the decomposition of the clique is equal to the sum of the claw and triangle decompositions. Thus, we can see that, by adding them, most of the details cancel out while the approximation graphs accumulate, giving evidence that we make emerge a large-scale structure. This can be more easily seen by just looking at the decomposition coefficients, which are also shown in the right panel. It can be seen that the scaling coefficients of the claw and triangle add to produce a new larger coefficient, while the largest-scale wavelet coefficients, $w^{(4)}_0$, cancel out, indicating that the details in this new graph will not longer be at a large scale but at finer scales (the self-loops in this case). Indeed, this cancellation effect between decomposition coefficients can be exploited to study the scales at which two graphs are different. Our next example illustrates this. 

\begin{figure}[t!]
\centering
  \includegraphics[width=0.48\textwidth]{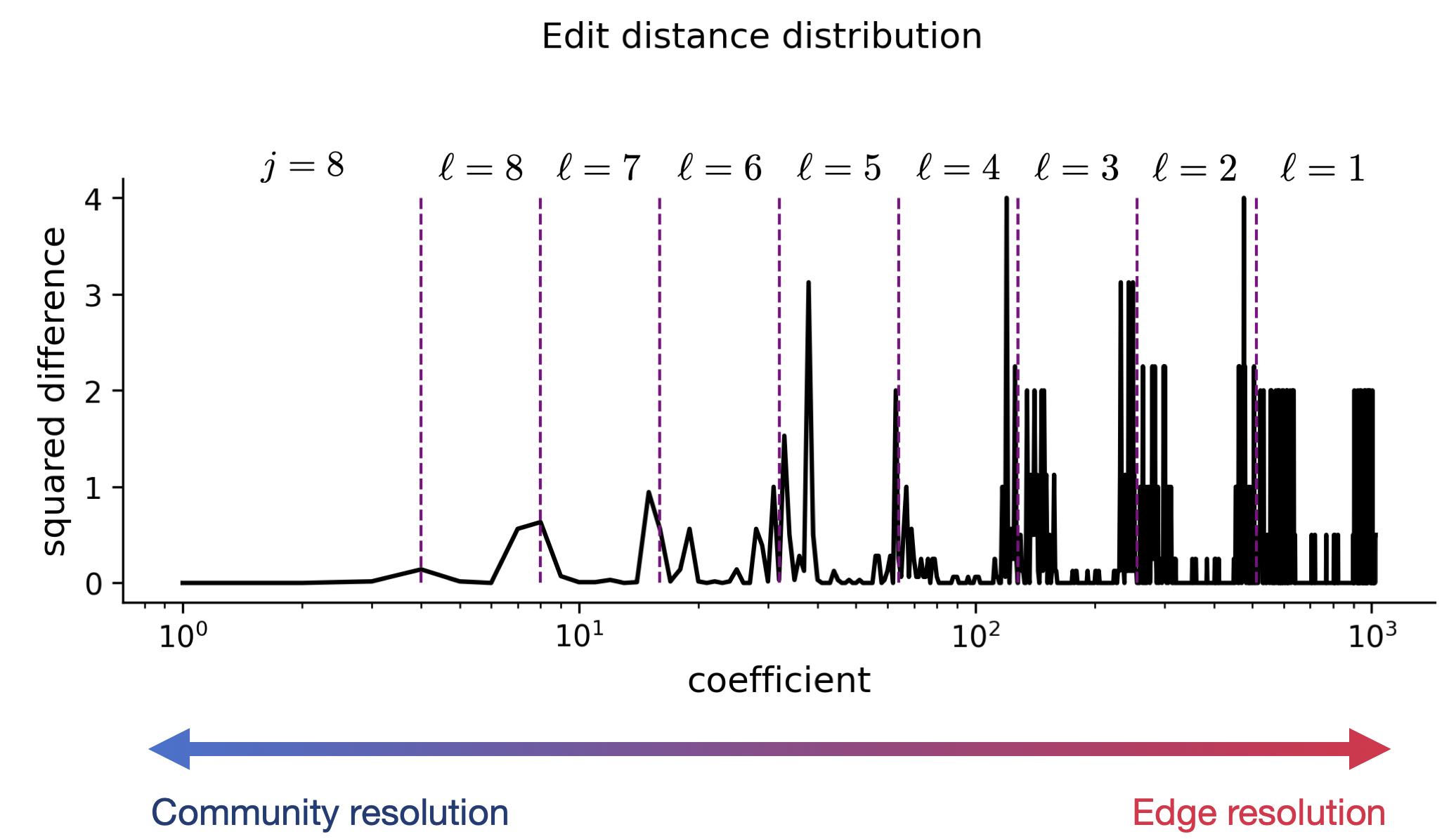}
  \caption{Distribution of the edit distance across decomposition coefficients for two realizations of the stochastic block model. Left-most coefficients refer to coarse-grain structures and right-most coefficients refer to detailed structures. Each coefficient captures a fraction of the edit distance. Coarse-grain structures contribute little to the edit distance, confirming the intuition that both realizations are similar at the community scale even thought their distance is large.}\label{Fig.sec.4.distance_distribution}
\end{figure}

{\bf Illustrating example (comparing graphs).}~One common way to assess the difference between two unweighted graphs is by means of their edit distance, counting their number of different edges. However, the edit distance is not entirely satisfactory in all situations, as we may have two graphs that can be considered structurally equal, yet their edit distance may still be very large. This is for instance the case of two realizations of a stochastic block model under equal parameters. In such case, the two realizations have very few common edges with high probability, meaning that their edit distance is large albeit the two graphs considered equal at the community level. 

Interestingly, our decomposition coefficients can help us to better spot the scales at which two graphs are equal or not while still keeping the interpretation in terms of edit distance. To show this, let us denote by $(s_{G_1})_k^{(j)}$ and $(w_{G_1})_k^{(j)}$ the decomposition coefficients of an unweighted graph $G_1$ and by $(s_{G_2})_k^{(j)}$ and $(w_{G_2})_k^{(j)}$ the coefficients of another unweighted graph $G_2$ (defined on the same relation-space). Then, we have that our decomposition coefficients satisfy the following property: 
\begin{multline}
\sum_k \left[ (s_{G_1})_k^{(j)} - (s_{G_2})_k^{(j)} \right]^2 + \\ \sum_{\ell \leq j} \sum_k \left[ (w_{G_1})_k^{(\ell)} - (w_{G_2})_k^{(\ell)} \right]^2 = edit(G_1, G_2)
\end{multline}
The proof of this property is given in Section \ref{Sec.Graph_Decomp_Embedding}. This equation indicates that our decomposition coefficients allow to study how the edit distance of two graphs distributes across resolution scales. Namely, since all the terms in the sum are strictly positive, then each term contributes with a fraction of the edit distance. This thus allows us to track the terms that contribute the most to the edit distance. Naturally, if we observe that the edit distance is mostly determined by the coefficients associated to a specific resolution level, then we can conclude that the differences between the graphs arise at such resolution level, also implying that the graphs are similar at all others resolution levels. 

In Figure \ref{Fig.sec.4.distance_distribution}, we show how these insights can be used to spot that two realizations of a stochastic block model are still equal at the community scale despite their large edit distance. For the figure, we generate two realizations of a block model with 2 blocks of 16 vertices each with within class probability $0.5$ and between class probability $0.01$. Then, we partition the relation-space so that, at resolution level $j = 8$, the sets $\mathcal{E}_k^{(8)}$, coincide with the community structure. We then apply our decomposition to such graphs based on the partitioning and display, in Figure \ref{Fig.sec.4.distance_distribution}, the squared difference of the coefficients. As it can be seen, the coefficients associated to large-scale structures contribute little to the edit distance, contrary to the detailed structures. This is thus consistent with our intuition that both graphs are equal at the community scale and that one needs to go to a finer scale in order to differentiate one from the other. In sum, the previous examples illustrate that our decomposition provides a useful tool to analyze the structural information of graphs, given that we appropriately choose our basis elements. In the next subsection, we investigate this crucial subject. 

\subsection{Partitioning of the relation-space}\label{Sec.Graph_Decomp_Partitioning}

While the derivations of the previous sub-section show that a multi-resolution analysis of graphs is possible, the crucial step of how to partition the relation-space is not addressed. This subsection investigates this point, for which we develop two techniques tailored for two common scenarios: (i) the case of graphs with community-like structures; and (ii) the case of activity graphs arising from fixed infrastructures. Independently of the scenario, we recall that our decomposition works best when the functions supported on the sets $\mathcal{E}^{(j)}_k$ are constant. Thus, our techniques essentially aim to find sets $\mathcal{E}^{(j)}_k$ of spatially close elements that mostly consist of active or inactive edges in the graph under analysis. 

{\bf SVD-based partitioning.}~When the graph to study has a community structure, a natural and adaptive approach to find the sets $\mathcal{E}^{(j)}_k$ consists in looking for rank-1 patterns of the adjacency matrix. The rationale is the following: since rank-1 patterns correspond to regions of the matrix that have similar state, then we can use such regions as our sets $\mathcal{E}^{(j)}$. Matrix factorization is the standard technique to find rank-1 patterns in a matrix. Therefore, we propose a methodology based on the second largest singular vector of the adjacency matrix to construct the sets $\mathcal{E}^{(j)}_k$. We stress that even though matrix factorization is a non-linear procedure, this done to fix the basis used to analyze the entire graph sequence, thus it does not impact our link stream framework. 

\begin{figure}[t]
\centering
  \includegraphics[width=0.48\textwidth]{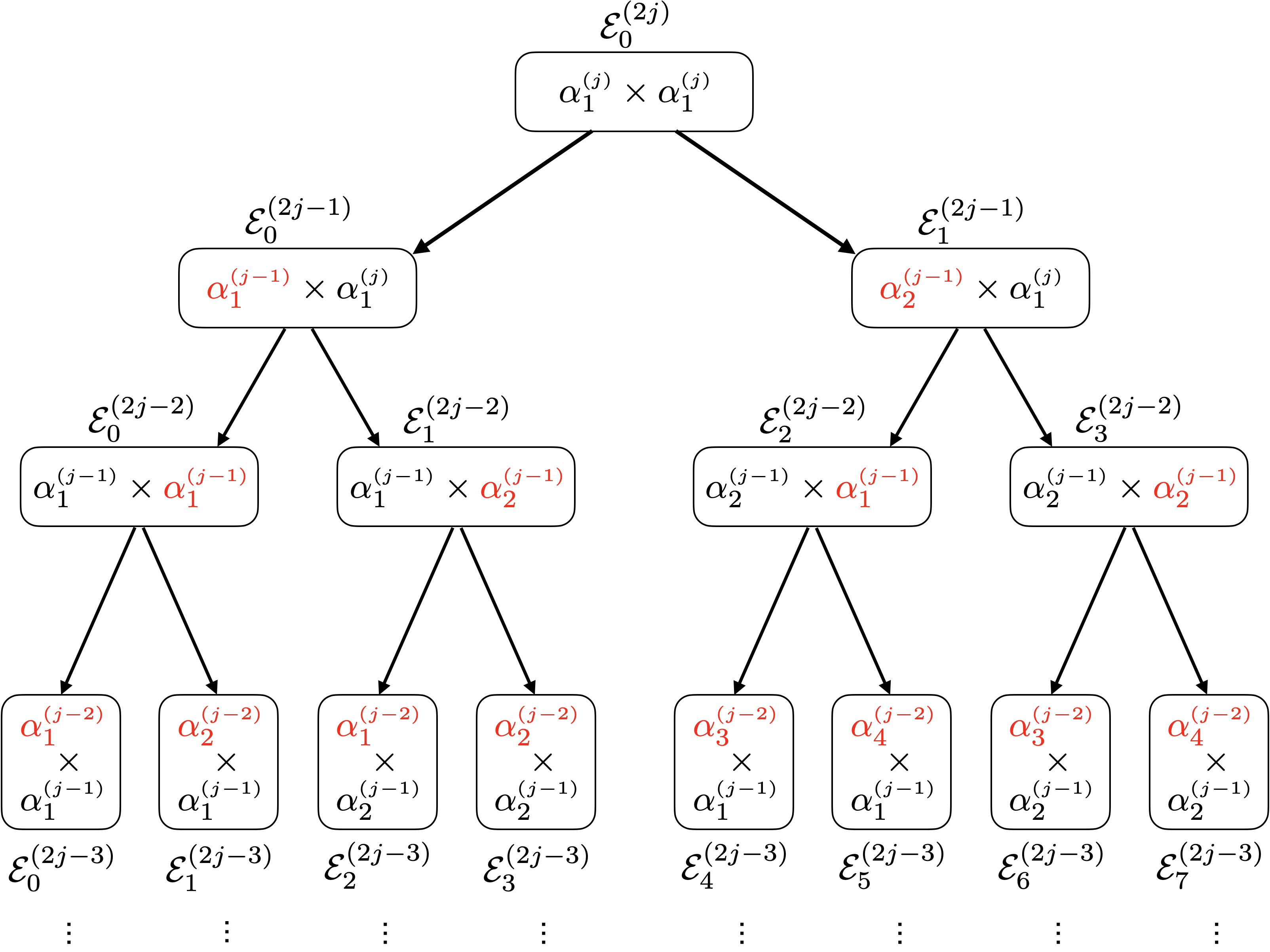}
  \caption{SVD-based procedure to partition the relation-space. The root node contains the entire relation-space which is recursively partitioned based on the splits found by the SVD procedure.}\label{Fig.sec.4.tree}
\end{figure}

Our partitioning algorithm works as follows. We first assign the set $\alpha^{(\log_2(N))}_1 = V$. Then, we recursively partition this set according to the following rule. For the set $\alpha^{(j)}_k$, we take the submatrix (of the adjacency matrix) of rows that are indexed by $\alpha^{(j)}_k$ (we keep all the columns). Then, we extract the second largest left singular vector of such sub-matrix and sort it in descending order. The elements of $\alpha^{(j)}_k$ associated to the top half entries of the singular vector form the set $\alpha^{(j - 1)}_{2k - 1}$ and the remainder form the set $\alpha^{(j - 1)}_{2k}$. We iterate this procedure until the sets $\alpha^{(0)}_k$ are singletons. For example, if $\alpha^{(2)}_1 = \{v_2, v_8, v_{12}, v_{20}\}$, then our procedure selects the rows associated to $v_2, v_8, v_{12}, v_{20}$ and forms a flat sub-matrix of size $4 \times |V|$. Assuming that the two largest entries of the left singular vector of this matrix are the ones associated to $v_2$ and $v_{20}$, then the procedure sets $\alpha^{(1)}_1 = \{v_2, v_{20}\}$ and $\alpha^{(1)}_2 = \{ v_8, v_{12} \}$. Based on the retrieved sets, we construct the motifs $\mathcal{E}_k^{(j)}$ according to the tree diagram shown in Figure \ref{Fig.sec.4.tree}. Essentially, a node contains a set $\mathcal{E}_k^{(j)}$ and its child nodes contain its division at the next resolution level. For odd levels, the division is done by separating edges according to their origin vertex, while for even levels the separation is done according to the destination vertex. For both cases, the sets $\alpha_k^{(j)}$ rule the splitting.

Interestingly, it is not necessary to build the tree and basis functions in practice in order to compute the coefficients of our decomposition. It suffices to know the sets $\alpha_k^{(0)}$ for all $k$. Namely, notice that the set $\mathcal{E}_i^{(0)}$, which contains a relation, implies a mapping of such relation to the integer $i$. We thus have that the leaf nodes of the tree define a hierarchy-preserving mapping of the relation-space to the interval $[1, M]$. This is, all the elements of the root node $\mathcal{E}_0^{(j_{max})}$ get mapped to the entire interval $[1, M]$, while those of the left child $\mathcal{E}_0^{(j_{max}-1)}$ get mapped to $[1, M/2]$ and those of the right child to $[M/2 +1, M]$, and so on. The implication of this result is that the scaling and wavelet functions defined over the sets $\mathcal{E}_{k}^{(j)}$, when mapped based on this rule, coincide with the classical Haar scaling and wavelet functions for time series defined on the interval $[1, M]$. Thus, we can transform our graph problem into a time series problem, which carries one main advantage: there exist fast implementations of the Haar wavelet transform for time series that avoid computing the basis elements and that use simple filter-banks instead, achieving time and space complexity of $\mathcal{O}(M)$. 

Notably, we can also avoid the construction of the tree as it is possible to find an analytic function that returns the position that a given relation has on the leaf nodes of the tree. To see this, notice that our SVD-based partitioning procedure implies a mapping of each vertex $u \in \alpha^{(0)}_k$ to a new unique index $k$. Then, we have that a relation $(u, v)$, relabelled as $(x, y)$, has a position in the leaf nodes determined by the following recursive function
\begin{equation}\label{eq.mapping}
z(x, y) = 
\begin{cases}
p^2 + z(x, y-p) & x \leq p ~~\text{and}~~y > p \\
2p^2 + z(x-p, y) & x > p ~~\text{and}~~y \leq p \\
3p^2 + z(x-p, y-p) & x > p ~~\text{and}~~y > p.
\end{cases}
\end{equation}
where $p$ is the previous integer to $max(x, y)$ that is a power
of 2. Based on this function, we can directly recover the index $i$ of the set $\mathcal{E}_i^{(0)}$ that contains $(u,v)$, implying that we can directly compute our decomposition coefficients by feeding the edges of the graph to this function and efficiently analyzing the resulting time series with a classical Haar filter bank. 

{\bf BFS-based partitioning}.
We now propose another adaptive procedure to partition the relation-space. This procedure is tailored for situations in which the graph models activity on a fixed infrastructure. For instance, in a road network, the graph can model if a road is being used, or in a wired computer network, the graph can model if a computer is communicating to another. In such situations, due to the fixed nature of the underlying infrastructure, some edges are impossible to appear in the activity graphs due to absence of physical connections between vertices. Thus, excluding them from the relation-space and hence from the motifs $\mathcal{E}_k^{(j)}$ can help to construct more meaningful basis elements. We therefore focus here on the study of functions reduced to a domain $\mathcal{E}_{active} \subset \mathcal{E}$, where $\mathcal{E}_{active}$ represents the set of relations where activity may occur. Our goal remains to meaningfully partition $\mathcal{E}_{active}$ to build the sets $\mathcal{E}_k^{(j)}$. For our discussion, we assume that $|\mathcal{E}_{active}| = M'$ is still a power of two.

Since structures of interest in these graphs have a strong notion of spatial locality, we address the partitioning of $\mathcal{E}_{active}$ by means of the breath first search algorithm (BFS). Our procedure works as follows. We first set $\mathcal{E}^{(\log_{2}(M'))}_{0} = \mathcal{E}_{active}$. Then, we recursively partition this set according to the following rule: for the graph with edge-set $\mathcal{E}_{k}^{(j)}$, we pick a node (which can be done at random) and we run the BFS algorithm from such node. Once the algorithm has explored $| \mathcal{E}_{k}^{(j)} | / 2$ edges, we associate the explored edges to the set $\mathcal{E}_{2k}^{(j-1)}$ and the non-explored edges to the set $\mathcal{E}_{2k +1}^{(j-1)}$. We continue until the sets $\mathcal{E}_{k}^{(0)}$ are singletons. 

Similarly to the SVD-based case, the resulting sets $\mathcal{E}_k^{(j)}$ can be organized in a tree where the leaf nodes define a hierarchy-preserving mapping of the relation-space to the interval $[1, M']$. As explained above, this implies that we can transform our graph decomposition problem into the one of decomposing a time series, which can be efficiently done via filter banks that avoid the construction of the basis elements. Thus, while our BFS-based method still requires to construct the tree, this remains rather simple to do due to the efficiency of the BFS algorithm. 

\subsection{Interpretation as graph embedding}\label{Sec.Graph_Decomp_Embedding}

In this subsection, we show that our decomposition coefficients can be used as a linear graph embedding technique. This is an important property, as classical embedding methods, which are based on neural networks or matrix factorizations, cannot be used as the matrix $\mathbf{Q}$ in our framework due to their non-linearity. To see our decomposition as an embedding method, it simply suffices to arrange the scaling coefficients into a vector 
\begin{equation}
	\mathbf{s} = [s^{(j)}_0, s^{(j)}_1, \dots] 
\end{equation}
and the wavelet coefficients into another one 
\begin{equation}
\mathbf{w} = [w_0^{(\ell)}, w_1^{(\ell)}, \dots]
\end{equation}
to form a unique vector 
\begin{equation}
\mathbf{x} = [ \mathbf{s}, \mathbf{w} ]
\end{equation}
that represents the embedding of the graph into a euclidean space of dimension $M$. While it is clear that the coefficients are individually related to the structural information of the decomposed graph, it is less clear that the geometry of points in the space also reflects the structural information of the embedded graphs. Our next result demonstrates this.

\begin{lemma}\label{Lemma.sec.4.embedding_full}
Let $G_1(V, E_1, f_{G_1}), G_2(V, E_2, f_{G_2})$ denote two unweighted graphs and $\mathbf{x_1}$, $\mathbf{x_2}$ their embedded vectors, respectively, under the same dictionary. Then, we have that:
\begin{enumerate}
	\item $\|\mathbf{x_1}\|_2^2 = |E_1| $ \vspace{5pt}
	\item $\langle \mathbf{x_1}, \mathbf{x_2} \rangle = |E_1 \cap E_2| $ \vspace{5pt}
	\item $\| \mathbf{x_1} - \mathbf{x_2} \|_2^2 = edit(G_1, G_2)$. 
\end{enumerate}
\end{lemma}

The proof of the lemma is deferred to Appendix \ref{Proof_Lemma_embedding_full}. It says that, for an unweighted graph, the size of the graph is preserved in the length of its embedding vector. Additionally, it says that if we measure the projection of one vector onto another, the result is equal to the number of common edges between the graphs represented by such vectors. Thus, this implies that two graphs with no common edges are orthogonal in the embedding space. Lastly, the lemma says that the distance between points in the embedded space preserves the edit distance between the graphs represented by such points. Thus, two nearby points in the embedded space necessarily imply that their associated graphs have small edit distance. 

Yet, our last result also implies that two graphs that may be structurally similar (like two realization of the stochastic block model) are still embedded far away as long as they have large edit distance. This shows that our embedding method adopts a graph similarity measure that may be too strict in some situations. To amend this issue, we can exploit the fact that the scaling coefficients capture the best coarse-grain approximation of a graph. Thus, by omitting the wavelet coefficients, we can obtain a graph similarity measure that looks for similarities at the structure scale. Our next result confirms this intuition. For it, we first formalize the notion of structurally similar graphs. Then, we introduce a lemma that demonstrates the properties of an embedding restricted to the sub-space spanned by $\bf{s}$. 
\begin{definition}
Let $G_1(V, E_1,f_{G_1})$ and $G_2(V, E_2, f_{G_2})$ be two unweighted graphs and $\mathcal{E} = \sqcup_k \mathcal{E}^{(j)}_k$ be a partitioning of the relation-space at resolution level $j$. Then, $G_1$ and $G_2$ are said to be structurally equal, at this resolution level, if it holds that $|E_1 \cap \mathcal{E}^{(j)}_{k}| = |E_2 \cap \mathcal{E}^{(j)}_{k}|$, for all $k$. 
\end{definition}
\begin{lemma}\label{Lemma.sec.4.embedding_lowdim}
Let $C_1$ and $C_2$ denote two classes of structurally equal graphs at a resolution level $j$ and assume that $G_1(V, E_1, f_{G_1})$, $\tilde{G}_1(V, \tilde{E}_1, f_{\tilde{G}_1}) \in C_1$ and $G_2(V, E_2, f_{G_2})$, $\tilde{G}_2(V, \tilde{E}_2, f_{\tilde{G}_2}) \in C_2$ are chosen uniformly at random. If we denote the scaling vectors of $G_1$ and $G_2$ by $\mathbf{s}_1$ and $\mathbf{s}_2$, respectively, then we have that they satisfy:
\begin{enumerate}
\item $\| \mathbf{s}_1 \|_2^2 = \mathbb{E} \left[E_1 \cap \tilde{E}_1\right]$ \vspace{5pt}
\item $\langle \mathbf{s}_1, \mathbf{s}_2 \rangle = \mathbb{E} \left[E_1 \cap E_2 \right]$ \vspace{5pt}
\item $\!\begin{aligned}[t]
\| \mathbf{s}_1 - \mathbf{s}_2 \|_2^2  = \mathbb{E}\left[edit(G_1, G_2)\right] - \\ \mathbb{E}\left[ dist(G_1, \tilde{G}_1) + dist(G_2, \tilde{G}_2) \right]
                \end{aligned}$
\end{enumerate}
\end{lemma} 

The proof of Lemma \ref{Lemma.sec.4.embedding_lowdim} is deferred to Appendix \ref{proof_lemma_embedding_lowdim}. It says that our new embedding no longer reflects the structure of individual graphs but rather the statistical properties of classes of graphs. For instance, previously, the length of a vector in the embedded space reflected the number of edges in the graph that mapped to such vector. Now, the length of the vector counts the expected number of edges that the graph has in common with all the graphs that are structurally equal to it. Similarly, the inner product of two vectors now counts the expected intersection between two classes of graphs rather than between two individual graphs. Notice that the last property also indicates that distance between points in this new embedding space is zero as long as the graphs are structurally equal. This therefore allow us to embed structurally similar graphs to nearby points even though they may have large edit distance. 

Compared to classical embedding techniques, our graph decomposition, when seen as an embedding, provides various advantages: (i) it is linear; (ii) it allows to express embedding vectors in analytic form; (iii) it allows to link the geometry of the space to the structure of graphs; (iv) it can be inverted; (v) it allows to analyze graphs at different resolution levels; (vi) it allows to characterize the details lost when considering coarse-grain information; (vii) it can be very efficiently computed with a simple filter-bank.

\subsection{Filters for graphs}\label{Sec.Graph_Decomp_Filters}

Our decomposition allows to represent graphs in a new structural domain. This naturally opens the door to study operators by characterizing how they change the structural-domain representation of a graph. Here, we perform this structural-domain study of graph operators. In particular, we focus on operators that can be represented as simple structural filters: operators that essentially amount to amplify or attenuate the coefficients of our decomposition. In more precise terms, we formalize the notion of structural filter as follows. 

\begin{definition}

Let $f_{G}$ represent a graph that is decomposed according to Equation (\ref{eq.sec.4.decomp_main}). Then, the structural filtering of $f_{G}$ is defined as 
\begin{equation}\label{Eq.sec.4.FilterStructDomain}
\hat{f}_{G}(e) = \sum_{k} \sigma_k^{(j)} s^{(j)}_k \phi^{(j)}_k(e) + \sum_{\ell \leq j} \sum_{k} \nu_k^{(\ell)} w^{(\ell)}_k  \theta_k^{(\ell)}(e)
\end{equation}
where $\{\sigma_k^{(j)}\}$ and $\{\nu_k^{(\ell)}\}$ are called the coefficients of the filter. 
\end{definition}

Thus, a structural filter transforms a graph $f_{G}$ into another graph $\hat{f}_G$ by tuning the importance given to scaling and wavelet functions. Equation (\ref{Eq.sec.4.FilterStructDomain}) provides the structural-domain definition of a filter, however it leaves unclear what type of transformations expressed in the initial domain of relations correspond to such structural filters. To answer this question, let 
\begin{equation}
f_{G}(\mathcal{E}_k^{(j)}) := \sum_{e \in \mathcal{E}_k^{(j)}} f_G(e). 
\end{equation}
Then, after simple algebraic manipulations to (\ref{Eq.sec.4.FilterStructDomain}), it is possible to show that transformations of the following form to $f_{G}$ behave like structural filters: 

\begin{align} \label{Eq.sec.4.FilterEdgeDomain}
\nonumber	\hat{f}_{G}(e) = \frac{\sigma_{k}^{(j)}}{2^j} f_G(\mathcal{E}_k^{(j)}) &+   \sum_{\ell \in \mathbb{L}_1} \frac{\nu_k^{(\ell)}}{2^{\ell}}[f_G(\mathcal{E}_{2k}^{(\ell-1)}) - f_G(\mathcal{E}_{2k+1}^{(\ell-1)})] \\ & + \sum_{\ell \in \mathbb{L}_2} \frac{\nu_k^{(\ell)}}{2^\ell}[f_G(\mathcal{E}_{2k+1}^{(\ell-1)}) - f_G(\mathcal{E}_{2k}^{(\ell-1)})]
\end{align}
where $e \in \mathcal{E}_k^{(j)}$, $\mathbb{L}_1 = \{ \ell : e \in \mathcal{E}_{2k}^{(\ell-1)}\}$ and $\mathbb{L}_2 = \{ \ell : e \in \mathcal{E}_{2k+1}^{(\ell-1)} \}$. Therefore, transformations that take averages from coarse-grain information and differences at smaller scales correspond to filters. 

\begin{figure}[t]
\centering
  \includegraphics[width=0.48\textwidth]{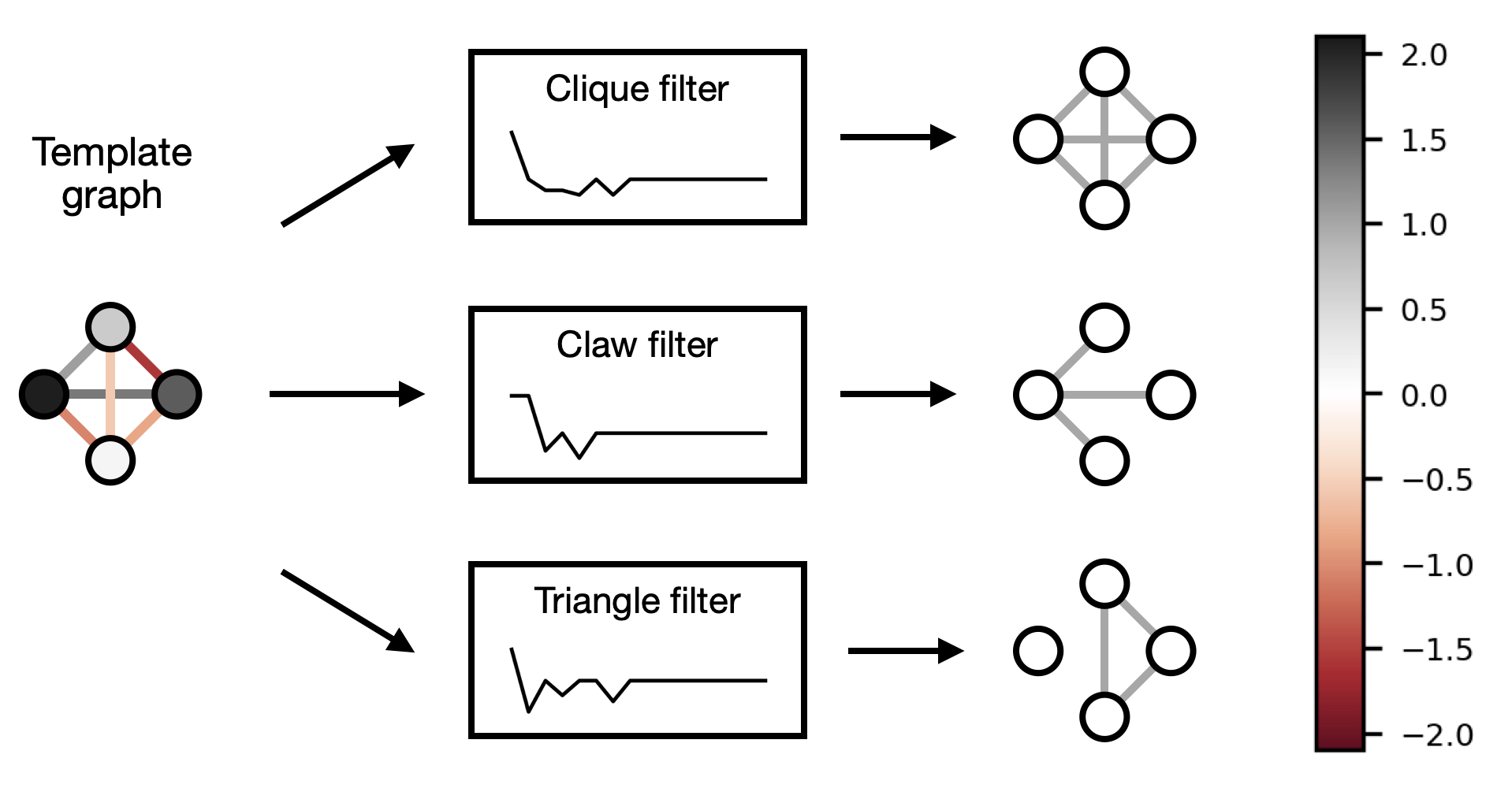}
  \caption{Illustration of the graph filtering process. A graph of four vertices that contains all structures with equal importance can be filtered out to produce any graph of four vertices with a desired structure.}\label{fig.sec.4.filtering_graphs}
\end{figure}

Structural filters open the door to interpret any graph as the result of filtering a pre-defined template graph that contains all possible structures with equal importance. Namely, consider a graph in which all its decomposition coefficients are equal to one. Then, this graph, which we call template graph, contains all the structures and can be transformed into any other graph by using filters that shape its structural-domain representation into a desired form. This is illustrated in Figure \ref{fig.sec.4.filtering_graphs}, where a template graph on four vertices is filtered to generate the claw, triangle and clique graphs. In terms of the graph embedding point-of-view of our decomposition, filters can be seen as transformations of the embedding vector of a graph into the embedding vector of another graph. In particular, filters can map a vector into any other vector that lives in the same sub-space. Thus, they model transformations between graphs living in the sub-space. In our example, the template graph lives on the largest sub-space and hence any graph can be produced from it. 

{\bf Coarse-grain pass filters.}~From (\ref{Eq.sec.4.FilterStructDomain}), two particular cases of interest arise. Firstly, the case in which the filter is chosen to preserve the scaling coefficients and to suppress the wavelet ones. This is, where $\sigma_k^{(j)} = 1$ for all $k$ and $\nu_k^{(\ell)} = 0$, for all $k$ and $\ell$. We coin this as a coarse-grain pass filtering of $f_{G}$, which we denote $\hat{f}^{(c)}_{G}$. By revisiting Fig. \ref{Fig.sec.4.decomposition}, it can be clearly seen that $\hat{f}^{(c)}_{G}$ corresponds to the coarse-grain approximations shown in the right panel. From (\ref{Eq.sec.4.FilterEdgeDomain}), we trivially have that these graphs are given in closed form as
\begin{equation}\label{Eq.sec.4.coarse-grain-filter}
	\hat{f}^{(c)}_{G}(e) = \frac{ f_{G}(\mathcal{E}_k^{(j)}) }{2^j}
\end{equation}
An interesting application of coarse-grain pass filters is the generation of graphs that are structurally similar to an input unweighted graph. To see this, notice that $\hat{f}^{(c)}_G$ corresponds to a new graph where all edges $e \in \mathcal{E}_k^{(j)}$ have a constant value between $[0, 1]$. If this value is taken as the probability of success of a Bernoulli trial, then by drawing such trial for each $e \in \mathcal{E}_k^{(j)}$ and interpreting successes as the edges of a new graph, we obtain a resulting graph that, on expectation, is structurally equal to $f_{G}$. Indeed, this procedure can be seen as a generalization of the stochastic block model to blocks determined by the sets $\mathcal{E}_k^{(j)}$ and where the probabilities are given by the input graph. 

{\bf Detail pass filters.}~The second case of interest consists in choosing the filter to suppress the scaling coefficients while preserving the wavelet ones. This is, the case where $\sigma_k^{(j)} = 0$ for all $k$ and $\nu_k^{(\ell)} = 1$ for all $k$ and $\ell$. We coin this as a detail-pass filtering of $f_{G}$, which we denote as $\hat{f}^{(d)}_{G_t}$. Since this filter contains the information not captured by the coarse-grain one, it corresponds to the detailed structures shown in the right panel of Figure \ref{fig.sec.4.filtering_graphs}. From (\ref{Eq.sec.4.coarse-grain-filter}), we readily have that
\begin{equation}\label{Eq.sec.4.detail_filters}
	\hat{f}^{(d)}_{G}(e) = f_G(e) - \frac{f_G(\mathcal{E}_k^{(j)})}{2^j}
\end{equation}
where $e \in \mathcal{E}_k^{(j)}$. This equation indicates that a detail filter provides a measure of the local variability of $f_{G}$ as it compares how different the weight of edge $e$ is with respect to the mean weight of edges in $\mathcal{E}_k^{(j)}$. It therefore can be interpreted as the graph analog of the differentiation operator for time series. Namely, in time series, differentiation consists in comparing a signal sample at time $t$ with the signal sample at time $t-1$, due to the ordered nature of time. However, in graphs, there is not such notion of ordered edges and it makes more sense to define differentiation by comparing the weight of an edge with the other weights in its local vicinity. This is precisely what our detail filters do, for which we define
\begin{equation}
	\frac{d f_G}{d e}(e) := \hat{f}^{(d)}_G(e)
\end{equation}
A useful consequence of having a notion of graph differentiation is that it opens the door to investigate regularity measures. Regularity measures are a key concept in machine learning, as they allow to reduce the space of admissible solutions to functions satisfying certain regularity properties. Essentially, the intuition is that a highly regular or smooth function (hence of small regularity metric) can be described with just a few parameters. Thus, it is easy to reconstruct or guess by just knowing a few of its samples. From this perspective, classical machine learning algorithms look for functions that fit the known samples and that are as regular as possible. In mathematical terms, the regularity of functions is often defined as the squared norm of the derivative of the function. Thus, adapted to our setting, we can define the regularity of $f_{G}$ as
\begin{equation}\label{Sec.LGM.Eq.regularity}
reg( f_G ) = \sum_{e \in \mathcal{E}}  \left( \frac{d f_{G}}{d e}(e) \right)^2
\end{equation}
Our next result demonstrates that this regularity metric indeed quantifies the complexity of guessing $f_G$. For it, it measures, in terms of graph distances, how hard it is to recover an unweighted graph from just having knowledge of its number of edges.
\begin{lemma}\label{Lemma.sec.4.graph_regularity}
Let $G(V, E, f_G)$ be an unweighted graph and $G^*(V, E^*, f_{G^*})$ denote another unweighted graph selected uniformly at random from the class of structurally equal graphs to $G$. Then, we have that 
\begin{equation}
reg\left( f_{G} \right) = \mathbb{E} \left[ \text{dist}(G, G^*) \right].
\end{equation}
\end{lemma}
The proof of Lemma \ref{Lemma.sec.4.graph_regularity} is deferred to Appendix \ref{proof.lemma.graph_regularity}. It says that if we try to recover $G$ by generating a random graph $G^*$ that has the same structure (which can be done by knowing the number of active edges in each set $\mathcal{E}_k^{(j)}$), then this regularity metric quantifies the expected error. Notice that zero error is only attained for the cases in which the motifs $\mathcal{E}_k^{(j)}$ correspond to either empty or complete sub-graphs in $G$. This is because the classes that contain such graphs only have them as members, thus any random selection from such classes recovers them. On the other hand, the metric attains maximum value when each motif has $|\mathcal{E}_k^{(j)}|/2$ active edges in $G$. This can be intuitively verified from the fact that the class that contains such graphs is the largest possible, thus the probability of selecting a graph $G^*$ that is close to $G$ is the smallest possible.

\section{Link stream analysis}\label{Sec.Link-Stream-Analysis}
\label{Sec.LS_Decomp}
In this section, we combine our graph developments of Section \ref{Sec.Linear-Graph-Methods} with their signal processing counterpart. We do this via the link stream analysis framework presented in Section \ref{Sec.Matrix_Framework}. Firstly, we combine the proposed graph decomposition with classical signal decompositions, like the Fourier or wavelet transforms. We show that this results in a new basis for link streams that allow us to represent them in a frequency-structure domain. We then use the framework to combine frequency and structural filters, allowing us to filter-out specific frequency and structural information from a link stream. Notably, we show that interesting transformations to link streams, like aggregation or embedding, correspond to simple filters in our frequency-structure domain. We finish by showing that filters naturally make emerge a notion of regularity of link streams, which opens the door to do machine learning on them.

\subsection{Frequency-Structure representation of link streams} 
\begin{figure*}[t!]
\centering
  \includegraphics[width=1\textwidth]{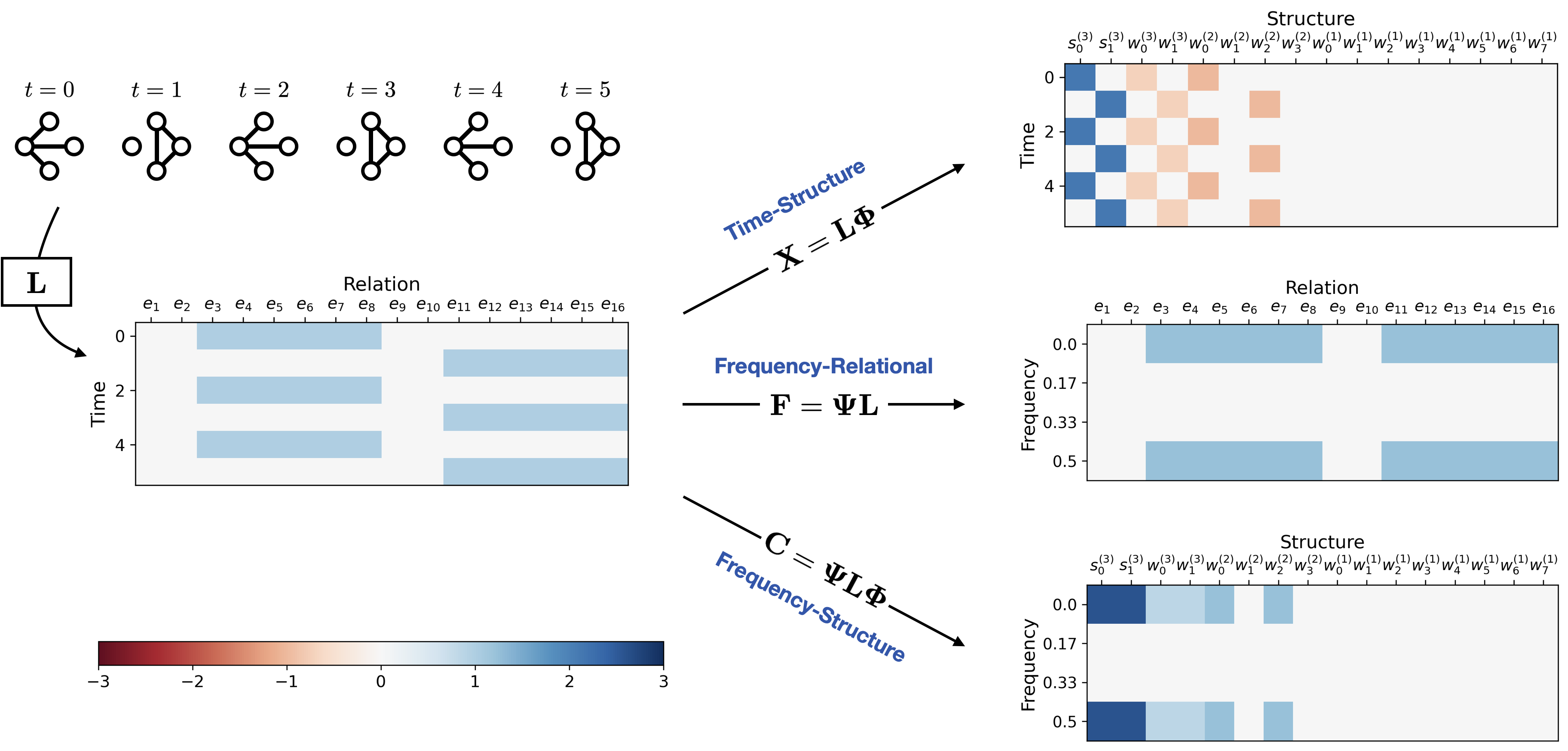}
  \caption{Illustration of the decomposition of link streams into their frequencies and structures. The oscillating claw and triangle link stream is encoded into a matrix (left panel). This matrix can therefore be projected into the graph basis (top-right), Fourier basis (center-right), or graph and Fourier basis (bottom-right). To graph basis is chosen as in Figure \ref{Fig.sec.4.decomposition}. The frequency-structure representation clearly reveals the importance the claw and triangle structures oscillating at frequency $1/2$.}\label{fig.sec.5.ls_decomposition}
\end{figure*}

We begin our analysis of link streams by combining our graph decomposition with classical time series ones. As a first step, we must fix a basis for graphs that is relevant to analyze the entire graph sequence. In the absence of extra information, the natural way to do this consists in aggregating all the link stream into a single graph. This graph then reveals all the regions of the relation-space where activity occurs and that are relevant to track. Based on such aggregated graph, we can then use one of our procedures from Section \ref{Sec.Graph_Decomp_Partitioning} (SVD-based or BFS-based) to partition the relation-space and fix the graph basis. As a second step, we encode the basis elements in a matrix $\mathbf{\Phi}$ where the rows contain, from top to bottom, the scaling functions and the wavelet functions: 
\begin{equation}
{\bf \Phi}^\top = [ \phi_0^{ (j)}, \phi_1^{(j)}, \dots, \theta_0^{ (\ell)},  \theta_1^{ (\ell)}, \dots]. 
\end{equation}
Then, by retaking the matrix representation of link streams $\mathbf{L}$, we notice that the coefficients of our decomposition can be simply computed, for the entire link stream, as 
\begin{equation}\label{seq.5.struct_decomp}
\mathbf{L} = \mathbf{X} \mathbf{\Phi},
\end{equation}
where the row $t$ of matrix $\mathbf{X}$ contains the decomposition coefficients associated to graph $G_t$. From (\ref{seq.5.struct_decomp}) we observe that (i) $\mathbf{L}$ can be exactly recovered from $\mathbf{X}$; and (ii) the rows and columns of $\mathbf{X}$ are indexed by time and structures, respectively. Therefore, $\mathbf{X}$ can be interpreted as a time-structure representation of $\mathbf{L}$. 

Similarly, we can represent $\mathbf{L}$ in a frequency-relational domain by defining the matrix ${\bf \Psi} = [\psi_1, \psi_2, \dots] $, where the columns are the atoms of a signal dictionary, like Fourier or wavelets. By focusing on the Fourier case, we can represent the frequency analysis of $\mathbf{L}$ as the simple matrix product 
\begin{equation}
{\bf L} = {\bf \Psi F}
\end{equation}
where column $j$ of matrix $\bf F$ contains the Fourier transform of $e_j(t)$. Since $\mathbf{F}$ is indexed by frequency and relations, and $\mathbf{L}$ can be entirely recovered from it, then we can consider $\mathbf{F}$ as a frequency-relational representation of $\mathbf{L}$. 

It can easily be seen that the matrices $\mathbf{\Psi}$ and $\mathbf{\Phi}$ take the role of $\mathbf{H}$ and $\mathbf{Q}$ in (\ref{Eq.sec.2.Joint-Analysis}), respectively, implying that they can be readily combined to express $\mathbf{L}$ in a frequency-structure domain. This is, by decomposing $\mathbf{F}$ into the graph dictionary (to extract the structural information from it), and $\mathbf{X}$ into the signal dictionary (to extract frequencies from it), we obtain in both cases the same matrix of coefficients $\mathbf{C}$ as:  
\begin{equation}\label{eq.sec.5.decomposition_ls}
\bf L = \Psi C \Phi. 
\end{equation}
The matrix $\mathbf{C}$ contains all the frequency and structure information of $\mathbf{L}$ as entry $\mathbf{C}_{uk}$ quantifies the importance of structure $k$ oscillating at frequency $u$ in $\mathbf{L}$. Equation (\ref{eq.sec.5.decomposition_ls}) constitutes our proposed decomposition for link streams. 

Two important points to highlight from (\ref{eq.sec.5.decomposition_ls}) are that (i) $\mathbf{\Psi}$ and $\mathbf{\Phi}$ are orthonormal matrices, therefore the inverse transformation is given by their complex conjugate, which we denote $\mathbf{\Psi}^\top$ and $\mathbf{\Phi}^\top$, respectively, for simplicity; and (ii) the expression can be rewritten as 
$ \mathbf{L} = \sum_{u,k} \mathbf{C}_{uk} \mathbf{Z}_{uk} $, 
where $\mathbf{Z}_{uk} = \psi_u \phi_k^\top$ is a link stream consisting of structure $k$ oscillating at frequency $u$, which is furthermore of unit-norm and orthonormal to any other $\mathbf{Z}_{u'k'}$ for $u' \neq u$ or $k' \neq k$, thus forming an orthonormal basis of link streams. 
\begin{figure*}[t!]
\centering
  \includegraphics[width=0.8\textwidth]{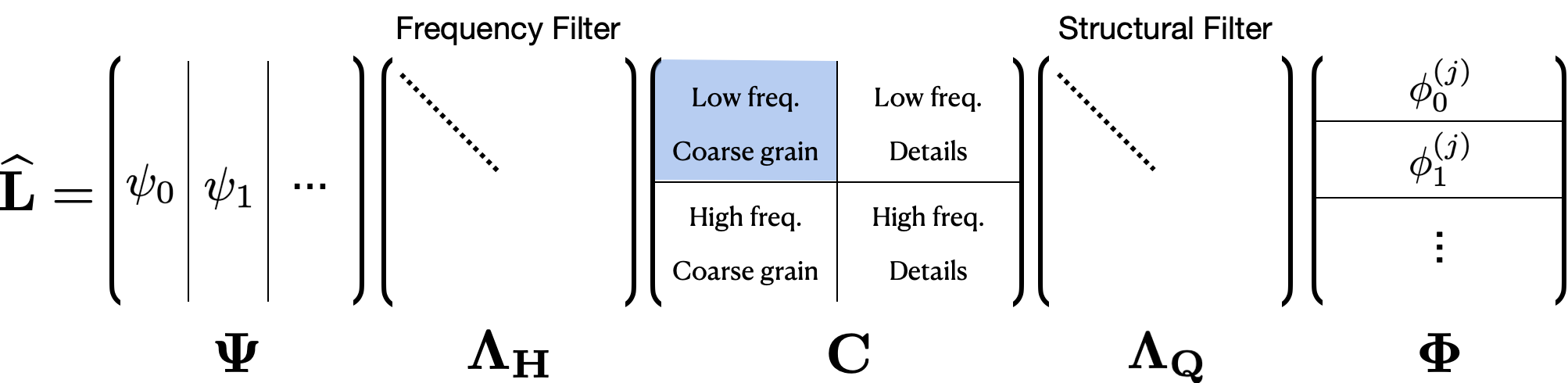}
  \caption{Illustration of the use of frequency and structural filters to recover specific information from the link stream. In the example, filters are designed to recover the backbone activity: coarse-grain structures with low frequencies.}\label{Fig.sec.5.ls_filters}
\end{figure*}

{\bf Illustrating example (oscillating link stream).}~
To better illustrate our link stream decomposition, Figure \ref{fig.sec.5.ls_decomposition} gives a practical example by revisiting the oscillating link stream introduced in Section \ref{Sec.Problem_Statement}. This link stream is composed of two clear structures (the claw and the triangle) oscillating at frequency 1/2 (i.e., they repeat every two samples). Therefore, we expect our decomposition to be able to reveal this information. 

We begin by encoding the link stream into the matrix $\mathbf{L}$ shown in the left-side of Figure \ref{fig.sec.5.ls_decomposition}. This matrix constitutes a time-relational representation of the data, as its entries encode the importance of a given relation at a given time. Since the rows of this matrix encode the graph sequence, we can obtain a time-structure representation by taking the product $\bf X = L \Phi{^\top}$. To do this, we must fix a basis $\mathbf{\Phi}$ which, for this example, is selected in the same way as in the example of Figure \ref{Fig.sec.4.decomposition}. The only difference is that we choose $j = 3$ since this resolution is more appropriate to study triangles and claws. Based on $\bf \Phi$, we compute $\mathbf{X}$ which is displayed at the top of the right-hand side of Figure \ref{fig.sec.5.ls_decomposition}. This representation permits to see that a claw and the triangle motifs, represented by the coefficients $s_{0}^{(3)}$ and $s_{1}^{(3)}$, are very important in the link stream, yet they do not appear simultaneously. 

On the other hand, since the columns of $\mathbf{L}$ encode the activation times of individual relations, then we can obtain a frequency-relational representation by taking the product $\mathbf{F} = \mathbf{\Psi}^\top \mathbf{L}$, where $\bf \Psi$ is the discrete Fourier transform matrix. For our example, we display the magnitude of $\mathbf{F}$ in the middle plot of the right-hand side of Figure \ref{fig.sec.5.ls_decomposition}. The figure clearly shows that interactions in $\mathbf{L}$ possess only two frequencies: $0$ and $1/2$. The $0$ frequency reflects the mean value of $e_k(t)$, which is clearly large due to the positivity of edge-weights. On the other hand, the $1/2$ frequency indicates that $e_k(t)$ displays the same behavior every two samples. 

In sum, $\mathbf{X}$ highlights the importance of the claw and triangle structures while $\mathbf{F}$ indicates the importance of the $0$ and $1/2$ frequencies. Yet, these disjoint representations are not fully satisfactory as they make it unclear which structures are related to which frequencies. To have a unique frequency-structure representation, we simply decompose $\mathbf{L}$ from both sides as $\mathbf{C} = \mathbf{\Psi}^\top \mathbf{L} \mathbf{\Phi}^\top $, for which the magnitude is shown in the bottom plot of the right-side of Figure \ref{fig.sec.5.ls_decomposition}. The figure clearly highlights that the most important structures are indeed the claw and the triangle and that both only oscillate at frequency $0$ (they have positively weighted edges) and frequency $1/2$ (they appear every two samples). Thus, this confirms that our decomposition effectively reveals the oscillating and structure nature of the link stream. 

\subsection{Filters in link streams}

We now show that frequency and structural filters can be combined in order to filter-out specific frequencies and structures from a link stream. We begin by adapting the structural filters to our matrix formalism. This is achieved by encoding the filter coefficients in the following diagonal matrix 
\begin{equation}
\mathbf{\Lambda_{Q}} = diag(\sigma_0^{(j)}, \sigma_1^{(j)},  \dots, \nu_0^{(\ell)}, \nu_1^{(\ell)}, \dots).
\end{equation}
This way, the filtering operation of the structural information of $\mathbf{L}$ can be represented as 
\begin{equation}
\widehat{\mathbf{X}} = \mathbf{X} \mathbf{\Lambda_Q}.
\end{equation}
To represent the filter in terms of $\mathbf{L}$, we have that  
\begin{align}\label{Eq.sec.5.Filter_LS_Domain}
\widehat{\mathbf{L}} &= \widehat{\mathbf{X}} \mathbf{\Phi} \\
&= \mathbf{X} \mathbf{\Lambda_Q} \mathbf{\Phi} \\
&= \mathbf{L} \mathbf{\Phi}^\top  \mathbf{\Lambda_Q} \mathbf{\Phi} \\
&= \mathbf{L}  \mathbf{Q}^{(\text{filt})} 
\end{align} 
where $ \mathbf{Q}^{(\text{filt})}  = \mathbf{\Phi}^\top  \mathbf{\Lambda_Q} \mathbf{\Phi} $. Hence, matrices that get diagonalized by $\bf \Phi$ constitute structural filters for link streams.

Similarly, we can adapt frequency filters to our matrix formalism by encoding the frequency response of the filter through the diagonal matrix 
\begin{equation}
\mathbf{\Lambda_{H}} = diag(\chi_0, \chi_1, \dots).
\end{equation}
where $\chi_i$ denotes the frequency response of the filter at frequency $i$. This way, the frequency filtering of $\mathbf{L}$ can be represented as
\begin{equation}
\widehat{\mathbf{F}} = \mathbf{\Lambda_H} \mathbf{F}.
\end{equation}
By similar derivations as in (\ref{Eq.sec.5.Filter_LS_Domain}), we have that the frequency filter can be expressed in terms of $\mathbf{L}$ as
\begin{equation}
\widehat{\mathbf{L}} = \mathbf{H}^{(\text{filt})}  \mathbf{L}
\end{equation}
where $\mathbf{H}^{(\text{filt})} = \mathbf{\Psi} \mathbf{\Lambda_H} \mathbf{\Psi}^\top$.  

We can then easily combine $\mathbf{H}^{(\text{filt})}$ and $\mathbf{Q}^{(\text{filt})}$ as
\begin{align}\label{Eq.sec.5.Filter_Combined_Filters}
\widehat{\mathbf{L}} &= \mathbf{H}^{(\text{filt})}  \mathbf{L} \mathbf{Q}^{(\text{filt})}  \\
& =   \mathbf{\Psi} \mathbf{\Lambda_H} \mathbf{\Psi}^\top \mathbf{L} \mathbf{\Phi}^\top  \mathbf{\Lambda_Q} \mathbf{\Phi} \\
& =  \mathbf{\Psi} \mathbf{\Lambda_H} \mathbf{C} \mathbf{\Lambda_Q} \mathbf{\Phi}.
\end{align}
From (\ref{Eq.sec.5.Filter_Combined_Filters}), we can see that $\mathbf{\Lambda_Q}$ suppresses the columns of $\mathbf{C}$ while $\mathbf{\Lambda_H}$ suppresses the rows. Thus, $\mathbf{\Lambda_Q}$ and $\mathbf{\Lambda_H}$ can be chosen to just let pass specific ranges of structures and frequencies. For example, they can be chosen to just let pass coarse grain structures that slowly oscillate as illustrated in Figure \ref{Fig.sec.5.ls_filters}. Our next examples explore in more detail the potential of these filters.

{\bf Illustrating example (aggregation and embedding).}~In Section \ref{Sec.Problem_Statement}, we stressed that the aggregation or embedding of link streams change their information in ways that are hard to characterize. Notably, our developments above allow us to show that aggregation or embedding can be seen as simple filters in our frequency-structure domain. To show this, let us notice that the $k$-sample aggregation operator can be represented by the matrix 
\begin{equation}
\mathbf{H}_{ij}^{(agg)} = 
\begin{cases}
1 & \text{if~} j - k < i \leq j \\
0 & \text{otherwise}
\end{cases}
\end{equation}
so that $\widehat{\mathbf{L}} = \mathbf{H}^{(agg)} \mathbf{L}$ denotes the aggregated link stream. $\mathbf{H}^{(agg)}$ is a circulant matrix and therefore is diagonalizable by the Fourier basis, meaning that it constitutes a frequency filter. 

\begin{figure}[t]
\centering
  \includegraphics[width=0.48\textwidth]{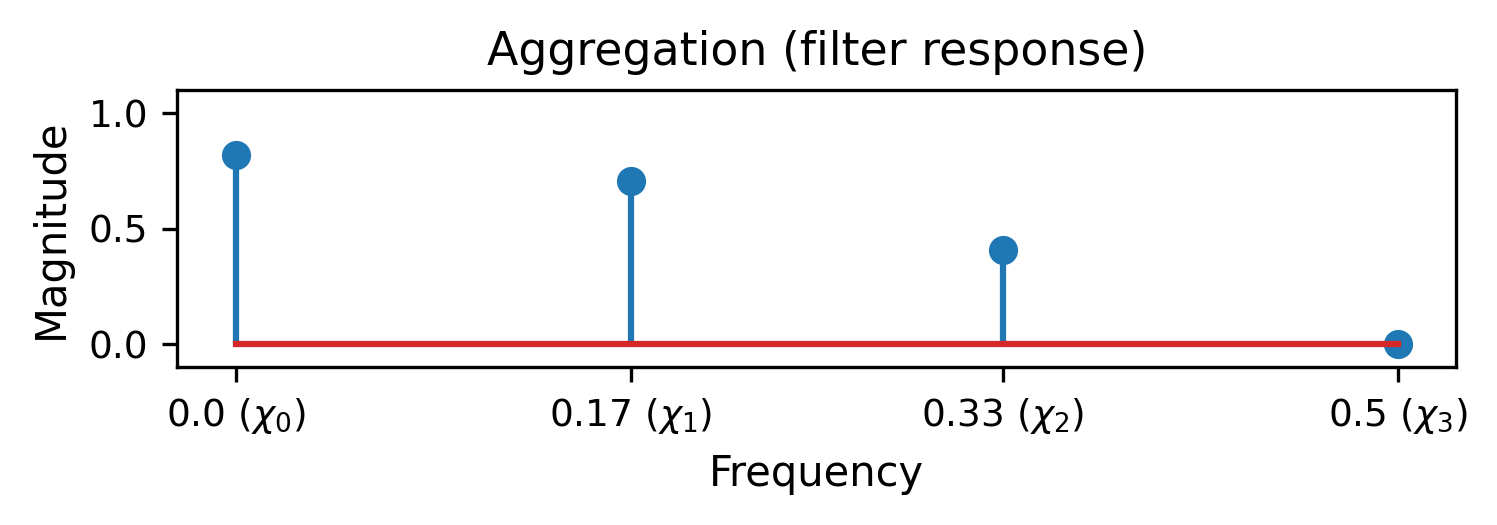}
  \includegraphics[width=0.48\textwidth]{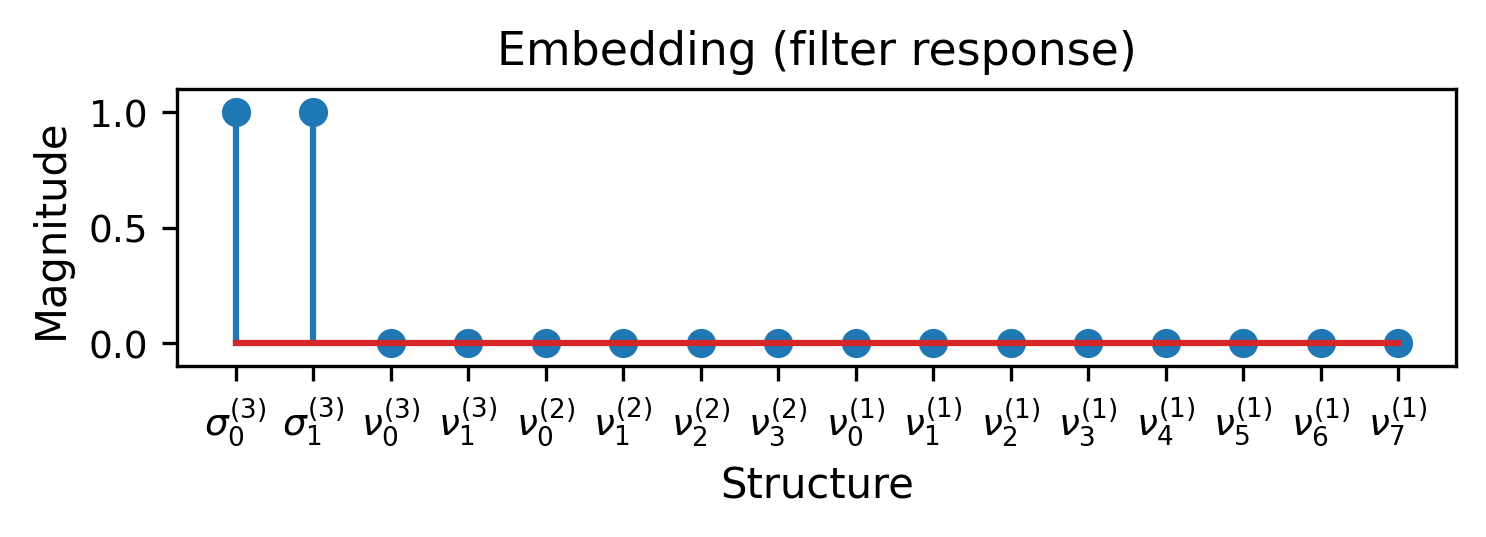}
  \caption{Aggregation and embedding processes modeled as filtering link streams. The two-sample aggregation of interactions corresponds to a low-pass filter in the frequency-domain. The embedding of graphs via the scaling coefficients corresponds to a coarse-grain filter in the structural-domain.}\label{Fig.sec.5.filter_embed_agg}
\end{figure}

In Figure \ref{Fig.sec.5.filter_embed_agg} (top), we display the frequency response of this filter for the $k=2$ case already used in our example of Section \ref{Sec.Problem_Statement}. As it can be seen, this filter lets the low frequencies pass while it entirely suppresses the information at frequency $1/2$. This implies that when it is applied to the oscillating link stream (see Figure \ref{fig.sec.5.ls_decomposition}), only the content at frequency $0$ is retained, resulting in a new link stream where the claw and triangle are now constant for all times. Since the simultaneous presence of the claw and triangle constitute a clique, then this explains why the 2-sample aggregation of our oscillating link stream results in a constant clique. 

Similarly, if we embed the link stream by applying our methodology proposed in Section \ref{Sec.Graph_Decomp_Embedding}, then we have that such procedure corresponds to applying a coarse-grain pass filter to $\mathbf{L}$. In Figure \ref{Fig.sec.5.filter_embed_agg} (bottom), we display the response of such filter, which only lets pass the structural coefficients. By applying this filter to our oscillating link stream (see Figure \ref{fig.sec.5.ls_decomposition}), then we can see that its effect is to just retain the coefficients associated to $s_{0}^{(3)}$ and $s_{1}^{(3)}$ for frequencies $0$ and $1/2$. Thus, this shows that the embedding method effectively maps the the link stream into two time series ($s_0^{(3)}(t)$ and $s_1^{(4)}(t)$) whose frequencies reflect the frequencies of the initial link stream. Of course, we may combine the aggregation and embedding filters and see that both amount to just retain the coarse-grain information at zero frequency. 

\begin{figure*}[t]
\centering
  \includegraphics[width=1\textwidth]{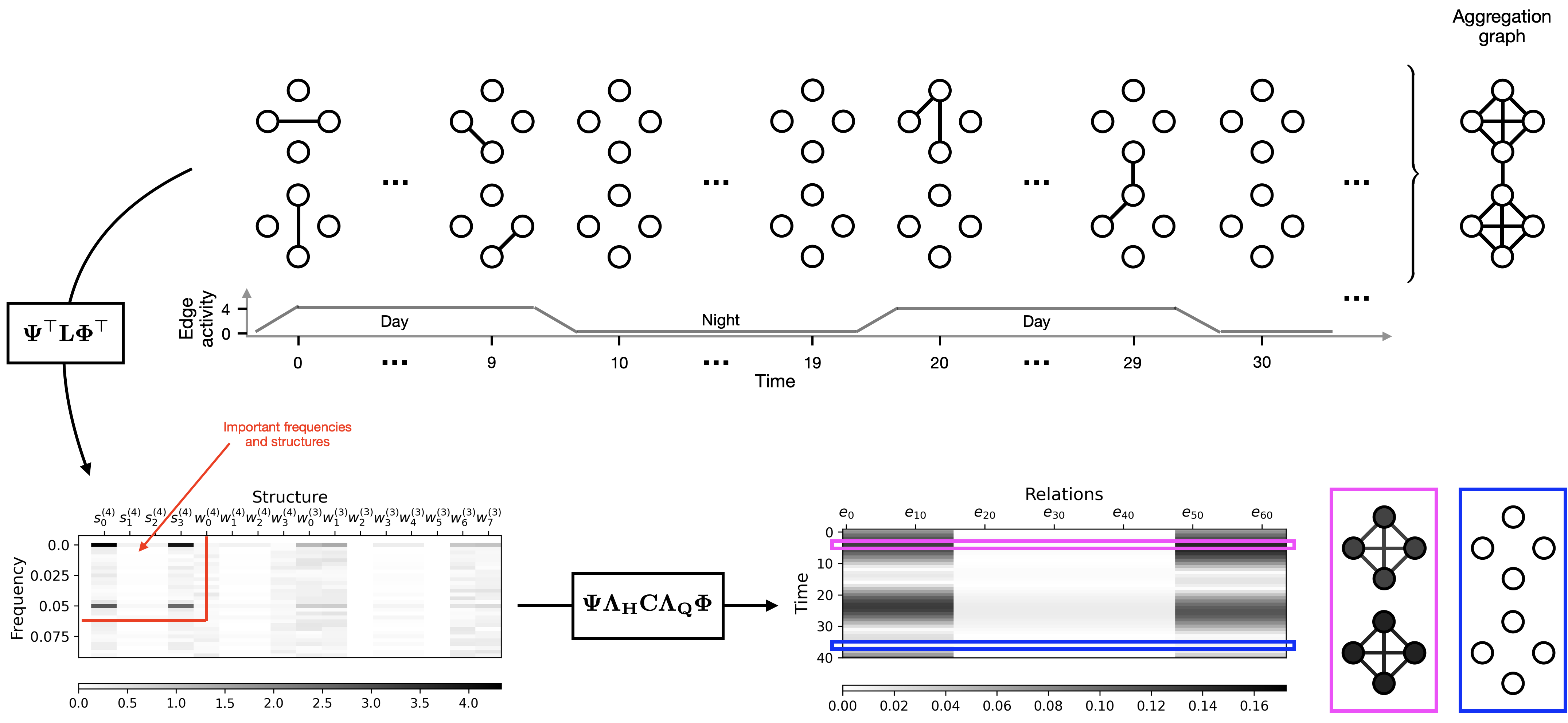}
  \caption{Example of the recovery of the backbone of a link stream. The raw link stream (top) consists of two-communities with day/night activity periods. The graphs of the sequence are too sparse to spot the community-like structure while the time-series of edges are too spiky to identify the day/night nature of the activity. The frequency-structure representation of the data (bottom left) reveals the importance of two communities with twenty-sample periodicity. Frequency and structural filters are therefore used to just retain the coefficients contained within the red-box. The resulting link stream (bottom right) reflects the data backbone: two fully active communities during the day and empty graphs during the night.}\label{Fig.sec.5.backbone}
\end{figure*}

{\bf Illustrating example (backbone of a link stream).}~In this example, our aim is twofold. Firstly, we aim to show that, even though a raw link stream may be extremely sparse, our decomposition allows to soundly reveal its characterizing frequencies and structures. Secondly, we aim to show that our filters can be used to retrieve the backbone of the link stream: its fundamental activity pattern. To show this, let us consider the example of Figure \ref{Fig.sec.5.backbone}, where it is shown (on the top) a link stream that reflects a typical communication pattern: two communities whose members sporadically communicate during daytime and are inactive during night-time. The challenge is that spotting this communication pattern from the raw data is a difficult task. Namely, the graphs from the sequence are too sparse to infer the community structure and the edge-time series are too spiky to infer the day-night periodicity. Normally, to see the pattern it is necessary to take aggregates of the link stream, which are unsatisfactory as they involve loss of information and different patterns may result in equal aggregates. 

Notably, our decomposition is able to effectively spot the pattern from the raw data. To show this, we employ our SVD-based partitioning procedure and report the magnitude of the frequency-structure coefficients of the link stream in the lower-left side of the figure. As it can be seen, there are four coefficients that contain most of the information: $s_{0}^{(4)}$ and $s_{3}^{(4)}$ both with frequency $0$ and $1/20$. These coefficients effectively indicate the presence of two communities whose activity periodically repeats every twenty samples. Interestingly, having four significantly large coefficients and many others with very little magnitude allow us to give a new interpretation of our link stream: it consists of two oscillating communities (large coefficients) plus details (small coefficients). We can therefore just retain the large coefficients in order to only preserve the non-detailed version of the link stream. This is what we call its backbone. To do this, we simply employ our frequency and structural filters to suppress the coefficients outside the red box shown in the plot. The resulting link stream is displayed in the lower-right part of Figure \ref{Fig.sec.5.backbone}. As it can be seen, it recovers the periods of large activity during daytime and periods of no activity during night-time. Moreover, the large activity periods consist of graphs with the two community structure. Thus, this effectively reflects the backbone of the original link stream. 

{\bf Illustrating example (regularity of link streams).}~In our previous example, we showed that low-pass frequency and coarse-grain structure filters allow to recover the backbone of the link stream. In this example, our aim is to show that high-pass frequency filters and detail structure filters allow to define a notion of regularity for link streams that paves the way to do machine learning on them. We recall that regularity metrics aim to measure the variability of a function, which is a good indicator of how hard it is to interpolate (predict), or to encode with a few parameters (compress). Since the link stream is a two-dimensional function of time and relations, then it is natural to define its regularity by taking its differentiation with respect to time and with respect to relations. To do this, let us notice that the time differentiation operator can be represented by the matrix 
\begin{equation}
\mathbf{H}^{(\text{diff})}_{uv} =
\begin{cases}
1 & \text{~if~} u = v \\
-1& \text{~if~} u = v + 1 \\
0 & \text{~otherwise},
\end{cases} 
\end{equation} 
so that the differentiation of $\mathbf{L}$ with respect to time can be simply expressed as 
\begin{equation}
\frac{\partial \mathbf{L}}{\partial t} = \mathbf{H}^{(\text{diff})} \mathbf{L}.
\end{equation}
Interestingly, $\mathbf{H}^{(\text{diff})}$ is a circulant matrix and therefore is diagonalizable by the Fourier basis, meaning that it constitutes a frequency filter. In particular, it is a high-pass frequency filter as it measures differences between successive time samples. Concerning the relational dimension, we showed in Section \ref{Sec.Graph_Decomp_Filters} that detail pass structural filters admit an interpretation as the graph analog of the differentiation operator. Therefore, we employ them to compute the differentiation of $\mathbf{L}$ with respect to the relational axis. To do this, notice that detail pass filters can be represented, in the time-relational domain, by the matrix
\begin{equation}
\mathbf{Q}^{(\text{diff})} = \mathbb{I} - \left( \frac{1}{2^j} \sum_{k}   \mathbbm{1}_{\mathcal{E}_k^{(j)}} \mathbbm{1}_{\mathcal{E}_k^{(j)}}^\top  \right),
\end{equation}
where $\mathbbm{1}_{\mathcal{E}_k^{(j)}}$ denotes the indicator vector of $\mathcal{E}_k^{(j)}$. Thus, differentiation of $\mathbf{L}$ with respect to relations can be expressed as
\begin{equation}
\frac{\partial \mathbf{L}}{\partial e} = \mathbf{L} \mathbf{Q}^{(\text{diff})}.
\end{equation}
Based on these definitions, we can then define the regularity of $\mathbf{L}$ along the relational axis as
$ reg_{e}\left(\mathbf{L}\right) = \left\lVert \frac{\partial \mathbf{L}}{\partial e} \right\lVert_F^2 $
and along the temporal axis as $ reg_{t}\left(\mathbf{L}\right) = \left\lVert \frac{\partial \mathbf{L}}{\partial t} \right\lVert_F^2 $, which can then be combined in order to obtain a measure of the total of variation of the link stream as
\begin{equation}
reg(\mathbf{L}) = reg_{t}\left(\mathbf{L}\right) + reg_{e}\left(\mathbf{L}\right). 
\end{equation}
To show that our regularity metric effectively measures the complexity of the link stream, our next result shows that, for unweighted link streams, the time regularity term measures the number of edge state changes over time and the relational regularity term measures the expected error when approximating the link stream by another random one that is structurally equal. 

\begin{lemma}\label{Lemma.sec.5.ls_regularity}
Let $\mathbf{L}$ be an unweighted link stream. Then, we have that $reg_{t}\left(\mathbf{L}\right) = \sum_{t} edit\left(G_t, G_{t-1}\right)$ and that $reg_{e}\left(\mathbf{L}\right) = \mathbb{E}\left[ \sum_{t} dist\left(G_t, G^*_t\right)\right]$, where $G_t^*$ is a graph drawn at random from the class of graphs structurally equal to $G_t$.
\end{lemma}

The proof of Lemma \ref{Lemma.sec.5.ls_regularity} is given in Appendix \ref{proof.lemma.ls_regularity}. From the lemma, it can be seen that zero regularity is only attained in the case where the following two criteria are met: the link stream does not evolve at all and the sequence is formed graphs where the sets $\mathcal{E}_k^{(j)}$ are either empty or complete. Therefore, only trivial link streams admit zero regularity. On the other hand, maximum regularity is attained when the following two criteria are satisfied: two successive graphs never intersect and the structures $\mathcal{E}_k^{(j)}$ have half of its edges active at all times. In sum, our regularity metric measures the slightest evolution and the slightest inactive edges in the structures of $\mathbf{L}$. 

Clearly, the above time regularity term measuring individual edge changes may be too strict for applications involving a sequence of structurally similar graphs: like a sequence of realizations from a stochastic block model which may consider regular due to its stable community structure, yet the above metric assigns a large regularity value. To address such situations, we can leverage our insights from Lemma \ref{Lemma.sec.4.embedding_lowdim} to propose a notion of time regularity at a larger structural resolution. For this, we notice that $\mathbf{X}$ can be seen as the concatenation of two matrices $\mathbf{X} = [\mathbf{S},  \mathbf{W}]$, where $\mathbf{S}$ contains the scaling coefficients and $\mathbf{W}$ the wavelet ones. Since the scaling coefficients act as an embedding method that maps structurally similar graphs to similar vectors, we can then define time differentiation at the structural resolution as $\frac{\partial \mathbf{S}}{\partial t} = \mathbf{H}^{(\text{diff})} \mathbf{S}$. Based on this, we can define a more relaxed temporal regularity term as $\text{reg}_t(\mathbf{S}) =  \left\lVert \frac{\partial \mathbf{S}}{\partial t} \right\lVert_F^2 $, which is zero as long as all the graphs from the sequence are structurally equal. 

We finish by stressing that these definitions open the door to do machine learning directly on link streams. To see this, recall that classical machine learning problem, in the supervised setting, refers to the problem of finding a target function which we partially know. Since the space of functions that fit the set of known values is potentially infinite, it is standard to reduce the space of admissible functions by imposing the a-priori that the target function should be highly regular (a property usually held by natural functions). Therefore, metrics that assess the regularity of functions are crucial in machine learning applications. In this vain, our regularity metric makes it possible to do machine learning in situations where the link stream, which is a 2-dimensional function, is partially known. For this problem and under the assumption that the link stream under search should be regular, our derivations above allow us to propose regularization problems of the form 
\begin{equation}
\text{argmin}_{\mathbf{L^*}}  \left\{ \| \mathbf{L} - \mathbf{L^*} \|_F^2 + reg(\mathbf{L^*}) \right\},
\end{equation} 
where the first term looks for a link stream that fits the known data and the second one penalizes irregular solutions. We leave the study of regularization problems on link streams as future work.

\section{Conclusion}
In this work, we presented a frequency-structure analysis of link streams. Our analysis is based on a novel linear matrix framework that represents link streams as simple matrices that unify their classical time series and graph interpretations. This representation then allows to process link streams by means of simple matrix products with matrices representing linear signal and graph operators. We showed that most signal processing operators can be readily adopted into our framework as a means to process time and we also developed a set of novel graph-based techniques to process the structural information. In particular, we developed a multi-resolution analysis for graphs and structural filters that allow us to spot and tune their structural information. These developments were possible by interpreting graphs as functions and then by adapting signal processing methods to process such functions. We showed that these results permit a novel representation of link streams in a frequency-structure domain that reveals the important structures and frequencies contained in it. Moreover, we showed that various interesting processing tasks can be seen as simple filters in this domain. In particular, our decomposition and filters open the door to extract features of better quality when searching for events of interest in link streams and also pave the way to do machine learning directly on them. 

\appendix
\subsection{Proof of Lemma \ref{Lemma.sec.4.embedding_full}}\label{Proof_Lemma_embedding_full}
\begin{proof}
Let $\mathbf{Q}$ denote the orthonormal matrix stacking the scaling and wavelet functions as its rows. Then, we have that $\mathbf{x_i}= f_{G_i}\mathbf{Q}^\top$. The proof of (2) follows from the fact that $\langle \mathbf{x_1} ,\mathbf{x_2} \rangle = (f_{G_1} \mathbf{Q}^\top) (f_{G_2} \mathbf{Q}^\top)^\top = f_{G_1}  \mathbf{Q}^\top \mathbf{Q} f_{G_2}^\top = f_{G_1} f_{G_2}^\top = | E_1 \cap E_2 |$, where we used the orthonormality of $\mathbf{Q}$ and the unweighted graph assumption. The proof of (1) follows as a particular case of (2) for $f_{G_2} = f_{G_1}$. To prove (3), notice that $\| f_{G_1} - f_{G_2} \|_2^2 = edit(G_1, G_2)$ for unweighted graphs. Then, by developing the left-hand side term and by using the orthonormality of $\mathbf{Q}$, we have that $\| f_{G_1} - f_{G_2} \|_2^2 = (f_{G_1} - f_{G_2})  (f_{G_1} - f_{G_2})^\top =  (\mathbf{x_1} - \mathbf{x_2}) \mathbf{Q} \mathbf{Q}^\top (\mathbf{x_1} - \mathbf{x_2})^\top = (\mathbf{x_1} - \mathbf{x_2})(\mathbf{x_1} - \mathbf{x_2})^\top$.

\end{proof}

\subsection{Proof of Lemma \ref{Lemma.sec.4.embedding_lowdim}}\label{proof_lemma_embedding_lowdim}

\begin{proof}
We begin by proving (2). To do it, let us notice that $\langle \mathbf{s}_1, \mathbf{s}_2 \rangle = \sum_k (s_1)^{(j)}_k (s_2)^{(j)}_k$. From the definition of structurally equal graphs, we have that $|E_1 \cap \mathcal{E}^{(j)}_k |  = |\tilde{E}_1 \cap \mathcal{E}^{(j)}_k |$ and that $|E_2 \cap \mathcal{E}^{(j)}_k |  = |\tilde{E}_2 \cap \mathcal{E}^{(j)}_k |$. This implies that, irrespectively of the sampled $G_1$ and $G_2$, the product $(s_1)^{(j)}_k (s_2)^{(j)}_k$ is invariant. By letting $|E_1 \cap \mathcal{E}^{(j)}_k |  = m_1$ and $|E_2 \cap \mathcal{E}^{(j)}_k |  = m_2$, we have that $(s_1)^{(j)}_k (s_2)^{(j)}_k = m_1 m_2 / 2^j$. Our goal now is to show that this quantity equals the expected number of common active tuples between $G_1$ and $G_2$ in the set $\mathcal{E}^{(j)}_k$. We do this by noticing that the intersection between $G_1$ and $G_2$, restricted to the tuples of the set $\mathcal{E}^{(j)}_k$, can be modeled as a sampling process without replacement. Namely, we can see the tuples of $\mathcal{E}^{(j)}_k$ as our total population consisting of $2^j$ possible tuples. From this population, the edges of $G_2$ embed $m_2$ of such tuples with the property of being active. Then, we take a sample of $m_1$ tuples from such population, where the sampled tuples are dictated by the edges of $G_1$. Our interest is to measure, from the sampled tuples, how many of them have being embedded with the property of being active as such tuples are active in both $G_1$ and $G_2$. This is, they are the number of common edges in both graphs. The process of sampling $m_1$ elements out of a population of $2^j$ elements where $m_2$ of them have some property is modeled by the hypergeometric distribution, where the expected number of sampled elements having the property is given as $m_1m_2 / 2^j$. Since the sets $\mathcal{E}^{(j)}_k$ are disjoint and expand the entire edge-space, then applying this argumentation to all $k$ proves (2). The prove of (1) follows as a particular case of (2) where $C_1 = C_2$. To prove (3), let us notice that $\| \mathbf{s}_1 - \mathbf{s}_2 \|_2^2 = \sum_k ((s_1)_k^{(j)} - (s_2)_k^{(j)})^2 $. By using the same assumptions as above, we have that
\begin{align}
((s_1)_k^{(j)} &- (s_2)_k^{(j)})^2 = \left(\frac{m_1 - m_2}{\sqrt{2^j}}\right)^2 \\
&= \frac{m_1^2}{2^j} + \frac{m_2^2}{2^j} - 2r \\
&= \frac{m_1^2}{2^j} + \frac{m_2^2}{2^j} - 2r + \gamma_1 + \gamma_2 - \gamma_1 - \gamma_2 \\
&= m_1 - r_{1,2} + m_2 - r_{1,2} - m_1 + r_{1,1} - m_2 + r_{2,2} \label{eq_prof_lemma2}
\end{align}
where $r_{i,j} = m_im_j/2^j$ and $\gamma_i = (1-\frac{m_i}{2^j})m_1$. The term $m_1 - r_{1,2}$ encodes for the number of edges in $G_1$ minus the expected number of edges in $G_1$ that are also in $G_2$ (restricted to $\mathcal{E}_k^{(j)}$). Therefore, if we let $G_1^{(k)}(V, E_1 \cap \mathcal{E}_k^{(j)})$ and $G_2^{(k)}(V, E_2 \cap \mathcal{E}_k^{(j)})$ denote the restrictions of $G_1$ and $G_2$ to $\mathcal{E}_{k}^{(j)}$, we have that 
\begin{equation}m_1 - r_{1,2} = \mathbb{E}[ dist(G_1^{(k)}, G_2^{(k)})]
\end{equation}
This means that (\ref{eq_prof_lemma2}) can be rewritten as
\begin{multline}
((s_1)_k^{(j)} - (s_2)_k^{(j)})^2 =  \mathbb{E}[dist(G_1^{(k)}, G_2s^{(k)}) + dist(G_2^{(k)}, G_1^{(k)}) \\
- dist(G_1^{(k)}, G_1^{(k)})- dist(G_2^{(k)}, G_2^{(k)})].
\end{multline}
The proof is finished after summing over $k$.   
\end{proof}

\subsection{Proof of Lemma \ref{Lemma.sec.4.graph_regularity}}\label{proof.lemma.graph_regularity}
For simplicity, let us denote $E \cap \mathcal{E}_k^{(j)} = E^* \cap \mathcal{E}_k^{(j)} = m_k$. Then, we have that 
\begin{align}
reg (f_{G}) &= \sum_{k} \sum_{e \in \mathcal{E}_k^{(j)}} \left(f_G(e) - \frac{m_k}{2^j} \right)^2 \\
	&= \sum_{k} \sum_{e \in \mathcal{E}_k^{(j)}} f_G(e)^2 + \frac{m_k^2}{2^{2j}} - \frac{2 m_k f_G(e)}{2^j} \\
	&= \sum_{k} m_k + \frac{m_k^3}{2^{2j}} - \frac{2 m_k^2}{2^j} + \frac{(2^j - m_k)m_k^2}{2^{2j}} \\
	&= \sum_{k} m_k - \frac{m_k^2}{2^j} \\
	&= \mathbb{E} [dist(G, G^*)]
\end{align}
where we last step follows from the same argumentation used in the proof of property (3) of Lemma \ref{Lemma.sec.4.embedding_lowdim}. 

\subsection{Proof of Lemma \ref{Lemma.sec.5.ls_regularity}}\label{proof.lemma.ls_regularity}

\begin{proof}
We have that $\| \frac{\partial \mathbf{L}}{\partial t} \|_F^2 = \sum_t \sum_e (f_{G_t}(e) - f_{G_{t-1}}(e))^2$. Due to the unweighted nature of the link stream, the term in the right-hand side is equal to $1$ if $f_{G_t}(e) \neq f_{G_{t-1}}(e)$ and $0$ otherwise. Therefore, $\sum_e (f_{G_t}(e) - f_{G_{t-1}}(e))^2 = edit(G_t , G_{t-1})$. The proof of the edge regularity term directly follows from Lemma \ref{Lemma.sec.4.graph_regularity} and the linearity of the expected value. 
\end{proof}

\bibliographystyle{ieeetr}
\bibliography{biblio_draft}
\end{document}